\documentclass[structabstract]{aa}  

\usepackage{natbib}
 \bibpunct{(}{)}{;}{a}{}{,} 
 
\usepackage{graphicx}

\usepackage{txfonts}
\def\micron{$\mu$m\space}

\def\micronno{$\mu$m}

\def\h2{H$_2$}
\def\Ms{M$_\odot$}
\def\ciis{[C\,{\sc ii}]}
\def\ciino{[C\,{\sc ii}]}
\def\hi{H\,{\sc i}\space}

\def\hii{H\,{\sc ii}\space}
\def\hiino{H\,{\sc ii}}
\def\cii{[C\,{\sc ii}]\space}
\def\13co{$^{13}$CO}
\def\c18o{C$^{18}$O}
\def\12co{$^{12}$CO}

\def\nii{[N\,{\sc ii}]\space}
\def\niii{[N\,{\sc iii}]\space}

\def\c+{C$^+$}
\def\Mss{$M_{\odot}$\space}
\def\h2{H$_2$}

\begin{document}

     \title{\cii emission from galactic nuclei in the presence of X-rays}
   
\titlerunning{\cii emission from  galactic nuclei in the presence of X-rays}

\authorrunning{Langer and Pineda}

   \author{W. D. Langer
              and
            J. L. Pineda
           }
          

   \institute{Jet Propulsion Laboratory, California Institute of Technology,
              4800 Oak Grove Drive, Pasadena, CA 91109-8099, USA\\
              \email{William.Langer@jpl.nasa.gov}
             }

   \date{Received 25 February 2015 / Accepted 27 May 2015}

 

\abstract
{The luminosity of \cii is used as a probe of the star formation rate in galaxies, but the correlation breaks down in some active galactic nuclei (AGNs).  Models of the \cii emission from galactic nuclei do not include the influence of X-rays on the carbon ionization balance, 
which may be a factor in reducing the \cii luminosity.}   
{We aim to determine the properties of the ionized carbon and its distribution among highly ionized states in the interstellar gas in galactic nuclei under the influence of X-ray sources. We calculate the \cii luminosity in galactic nuclei under the influence of bright 
sources of soft X-rays.} 
{We solve the balance equation of the ionization states of carbon as a function of X-ray flux, electron, atomic hydrogen, and molecular hydrogen density. These are input to models of \cii emission from the interstellar medium (ISM) in galactic nuclei  representing conditions in the Galactic Central Molecular Zone and a higher density AGN model. The behavior of the \cii luminosity is calculated as a function of the X-ray luminosity. 
We also solve the distribution of the ionization states of oxygen and nitrogen in highly ionized regions. }  
{We find that the dense warm ionized medium (WIM)  and dense photon dominated regions (PDRs) dominate the \cii emission when no X-rays are present.  The X-rays in galactic nuclei can affect strongly the C$^+$ abundance in the WIM, converting some fraction to C$^{2+}$ and higher ionization states and thus reducing its \cii luminosity.  For an X-ray luminosity $L$(X-ray) $\gtrsim$ 10$^{43}$ erg s$^{-1}$ the \cii luminosity can be suppressed by a factor of a few, and for very strong sources, $L$(X-ray) $>$10$^{44}$ erg s$^{-1}$ such as found for many AGNs, the \cii luminosity is significantly depressed. Comparison of the model with several extragalactic sources shows that the \cii to far-infrared ratio declines for  $L$(X-ray) $\gtrsim$10$^{43}$ erg s$^{-1}$, in reasonable agreement with our model.} {We conclude that X-rays can suppress the C$^+$ abundance and, therefore, the \cii luminosity of the ISM in active galactic nuclei with a large X-ray flux. The X-ray  flux can arise from a central massive accreting black hole and/or from many smaller discrete sources distributed throughout the nuclei.  We also find that the lower ionization states of nitrogen and oxygen are also suppressed at high X-ray fluxes in warm ionized gas.}

{} 

\maketitle



\section{Introduction}
\label{sec:introduction}

The fine structure line of ionized carbon, \ciino, at 158 \micron is widely used to trace the galactic star formation rate (SFR)  \citep{Stacey2010,DeLooze2011,DeLooze2014,Pineda2014}, as is the far-infrared dust emission \cite[e.g.][]{Malhotra1997,Luhman1998,Stacey2010,DeLooze2014}.  In principle the energy output of bright young stars is transferred to the gas and dust and reemitted either in the important far-infrared cooling lines such as \cii or as continuum dust emission.  
In extragalactic sources the luminosity of \cii is often correlated with the far-infrared dust emission thus reinforcing the suggestion that these trace the SFR.  
However there are some notable exceptions in which the luminosity of \cii is significantly suppressed with respect to the infrared luminosity. The relationship between the intensity of \cii and other star formation tracers appears to work in the Galactic disk \citep{Pineda2014} and most extragalactic sources \citep{DeLooze2014}, however it breaks down somewhat in many luminous infrared galaxies (LIRGs) \citep{Diaz-Santos2014} and ultra luminous infrared galaxies (ULIRGs) \citep{Malhotra1997,Luhman1998,Rigopoulou2014} and even more so in many active galactic nuclei (AGN) \cite[c.f.][]{Stacey2010,Sargsyan2014,Gullberg2015}. The assumption is that there is a deficit in \cii luminosity in these sources and in particular in their central regions \citep{Malhotra1997,Luhman1998,Rigopoulou2014,Gullberg2015}.

Several explanations have been offered for the decrease in the \cii to far-infrared luminosity \cite[e.g.][]{Nakagawa1995,Malhotra1997,Luhman1998,Stacey2010,Gullberg2015}, including: 1) the \cii emission is optically thick and saturated, if it arises from very dense photon dominated regions (PDRs), where the excitation temperature of the $^2P_{3/2}$--$^2P_{1/2}$ transition is thermalized and does not radiate as efficiently as the dust with increasing energy input; 2)  for high ratios of UV flux to density the dust grains are positively charged and are less efficient at heating the gas via photoejected electrons, thus the $^2P_{3/2}$ level is less populated \citep{Wolfire1990,Kaufman1998};  and, 3) a reduction in the column density of ionized carbon in the PDRs due to a soft ultraviolet radiation field that can heat the dust but not maintain a large layer of C$^+$. However, detailed models of the source of \cii are lacking and none of the proposed mechanisms appear to explain the reduction in all sources.

In this paper we propose that X-rays can also produce a deficiency in the \cii to far-infrared (FIR) luminosity ratio in the presence of a large X-ray flux at energies above $\sim$1 keV. Under these conditions, a significant fraction of carbon is in higher ionization states (C$^{2+}$, C$^{3+}$, etc.) in some of the regions responsible for \cii emission,  in particular the highly ionized ISM gas components.  We also propose that X-ray photoionization from strong X-ray sources can reduce the abundance of the lower ionization states of nitrogen and oxygen in the ionized gas and therefore the emission from their far-infrared fine-structure lines. For example, the \niii 57 \micron line is observed to be comparable to the \nii 122 \micron line in extragalactic sources \citep{Gracia-Carpio2011}. The abundance of N$^{++}$ will be comparable to or greater than that for N$^+$ because the critical density for exciting the N$^{++}$ fine-structure levels is much higher than that of those for N$^+$\citep{Madden2013}. Previous work on the effects of X-rays on chemistry have focused on the X-ray Dissociation Regions (XDRs) of molecular clouds \cite[e.g.][]{Langer1978,Krolik1983,Maloney1996,Meijerink2005} or YSOs  \citep{Bruderer2009} and protoplanetary disks \citep{Adamkovics2011}\footnote{There is also a body of work on the carbon ionization balance in very hot diffuse ionized gas due to electron collisional ionization of carbon \cite[e.g.][]{Gnat2007}.}.


Here we consider the effects of high X-ray fluxes on the interstellar gas environments that emit in \ciino. These include primarily the highly ionized gas, which is composed of several components including the warm ionized medium (WIM), Hot Ionized Medium (HIM), the dense ionized skins surrounding  clouds (DIS), and \hii regions.  A large X-ray flux will also affect the \cii emission from diffuse atomic hydrogen clouds and, to a much lesser extent, the photon dominated regions (PDRs) surrounding molecular clouds.   We  also consider the ionization balance of nitrogen and oxygen in the WIM. The emphasis in this paper is different from that in the XDR models \citep{Maloney1996,Meijerink2005} in that we do not evaluate the molecular abundances in the dense UV shielded molecular clouds, but focus instead on the abundance of the carbon ionization state in the major ISM components. Whether X-rays play a significant role in reducing the abundance of C$^+$  (and other gas tracers such as N$^+$, N$^{2+}$, O$^0$ and O$^{2+}$) in the ISM and, thus, \cii emission will depend on details of the distribution of X-ray sources, their  luminosity, and the properties of the ISM.  In addition, a high flux of X-rays can compete with UV heating resulting in an increase in the far-infrared (FIR) luminosity \citep{Voit1991b} and therefore further reducing the \cii to FIR ratio.

We will show that the X-ray flux can reduce the abundance of C$^+$ primarily in highly ionized regions and that, under the right conditions, will reduce the \cii emission significantly in galactic nuclei.  In the Galactic disk the WIM contributes about 20\% to 30\% of the \cii luminosity \citep{Pineda2013,Velusamy2014} using the scale heights derived from \cii \citep{Langer2014z,Velusamy2014}. In the Galactic Central Molecular Zone (CMZ)  the WIM densities are much higher and may contribute an even larger percentage of the \cii luminosity. We conclude that this mechanism is important in some active galactic environments with large X-ray luminosity and must be included along with the previously suggested mechanisms to explain the reduction in the \cii to far-IR luminosity in galaxies.

\begin{figure}
 \centering
            \includegraphics[width=9.3cm]{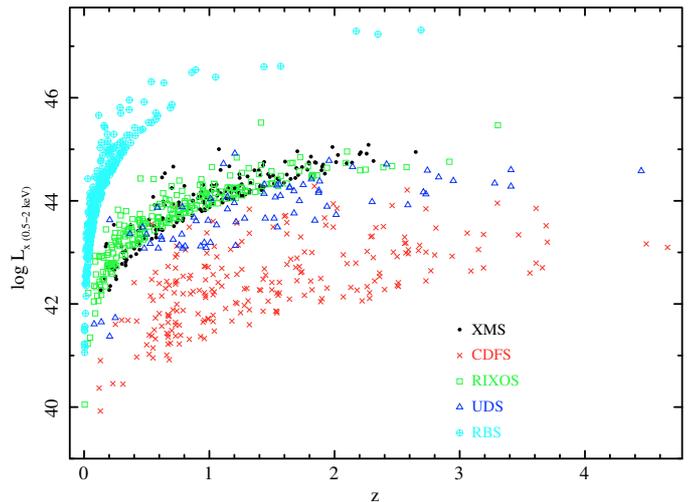}
      \caption{Soft X-ray (0.5 -- 2.0 keV) luminosity for a sample of 1009 AGNs as a function of red shift $z$ reproduced from Fig. 1 in \citet{Ebrero2009}. The insert key identifies the X-ray survey used to compile the data (see \citet{Ebrero2009} for their definitions).  The soft X-ray luminosities for this sample of AGNs range from $\sim$10$^{40}$ to $\sim$2$\times$10$^{47}$ erg s$^{-1}$.}
      \label{fig:fig1}
 \end{figure}

X-rays are an important radiation component of active galactic nuclei (AGNs) composing up to 40\% of of the total luminosity \citep{Mushotzky1993}.   \cite{Ebrero2009} summarize the soft (0.5 -- 2 keV) X-ray luminosities in  a large sample ($>$ 1000) of AGNs (reproduced in Fig.~\ref{fig:fig1}) which have luminosities in the range 10$^{40}$ erg s$^{-1}$ to 2$\times$10$^{47}$ erg s$^{-1}$. \cite{Iwasawa2011} in a recent study of Luminous Infrared Galaxies (LIRGs) found that 43 out of 44 sources  had X-rays luminosities ranging from $\sim$10$^{40}$ erg s$^{-1}$ to $\sim$10$^{43}$ erg s$^{-1}$  in the 0.5 to 10 keV range. The mean radius of the soft X-ray emission in their sample was 5.3 kpc and so represents emission throughout the LIRG, however  they found that the distribution of the hard X-ray emission is much more compact.  X-ray luminosities greater than 10$^{40}$ erg s$^{-1}$ represent an important ionization source that must be taken into account in understanding the conditions of the interstellar medium and the corresponding emission in far-infrared lines from ions.

Even in more quiescent galaxies X-rays are important in the nucleus.  For example,  the Galactic CMZ is observed to have extended diffuse X-ray emission as well as numerous X-ray point sources \citep{Muno2009}, including an accreting stellar black hole \citep{Heindl1993}.  In addition, the massive black hole near {\it Sgr A} is a potential X-ray source. It has been suggested that X-rays could provide the $>$ 10$^{-15}$ s$^{-1}$ ionization rate of hydrogen needed to explain the H$_3$$^+$ observations of \cite{Goto2014}. The cross section for ionization of heavy atoms by X-rays is orders of magnitude larger than hydrogen,  primarily due to K shell ionization at X-ray energies greater than a few hundred eV.  In addition, X-ray photoionization of hydrogen produces energetic photoejected electrons which produce secondary ionization of carbon \citep{Maloney1996,Meijerink2005}.  Thus, X-rays can play an important role in ionizing carbon and other metals. The penetration depth of 1 keV X-rays in the diffuse ionized gas in the CMZ  is a column density $N$(H$^+$) $>$ few$\times$10$^{22}$ cm$^{-2}$ and in the neutral gas, is $N$(H+2H$_2$) $\sim$ several $\times$10$^{21}$ cm$^{-2}$. Thus X-rays suffuse galactic nuclei and X-ray photoionization needs to be considered in the ionization balance of metals in the interstellar medium. 

Our paper is organized as follows.  In Section~\ref{sec:observations} we develop the reactions and ionization balance model, while in Section~\ref{sec:results} we present sample model calculations of the distribution of ionization states of C, N, and O. In Section~\ref{sec:discussion} we model the \cii luminosity as a function of X-ray luminosity, and discuss the consequences and implications of X-ray photoionization on the \cii emission from active galactic nuclei. In Section~\ref{sec:discussion} we also validate our  \cii emissivity model for galactic nuclei by calculating the luminosity from the CMZ and comparing it to observations.  Finally,  we summarize our results in Section~\ref{sec:summary}.



\section{X-ray ionization model for carbon}
\label{sec:observations}

In this section we develop the X-ray photoionization balance equations for the ionization states of carbon.  Multiply ionized carbon can be produced by L- and K-shell photoionization by X-rays.  K-shell photoionization dominates L-shell photoionization at energies above its threshold, $E_K \sim$ 0.3 -- 0.4 keV for carbon. Furthermore, these higher  energy X-rays penetrate further in the ISM than the low energy X-rays that dominate L-shell photoionization because the L-shell photoionization cross sections are large for hydrogen and carbon near their thresholds $\sim$ 11 to 15 eV.  We consider only X-rays with energies $\gtrsim$ 1 keV because they have large enough cross sections to influence the ionization balance but not so large that they cannot penetrate large column densities. 

\subsection{Carbon ion balance equations}

Here we describe the reduced set of reactions that dominate the ionization balance of carbon in the presence of X-rays.  The X-ray photoionization processes are:

\begin{equation}
h\nu_X + C^{j+} \rightarrow C^{(j+1)+} + e
\label{eqn:X_L_photo}
\end{equation}

\begin{equation}
h\nu_X + C^{j+} \rightarrow C^{(j+2)+} + 2e
\label{eqn:X_K_photo}
\end{equation}

\noindent where $j$ labels the initial ionization state, C$^j$, and  ranges from 0 to 5 for L-shell  ionization (Equation~\ref{eqn:X_L_photo}) and 0 to 4 for Auger K-shell ionization (Equation~\ref{eqn:X_K_photo}).  
The X-ray photoionization rate from C$^j$  is  designated $\zeta_{XL}^j$ and $\zeta_{XK}^j$ for L-shell and K-shell, respectively.   

The electron recombination reactions are,

\begin{equation}
C^{(j+1)+} + e \rightarrow C^{j+} + h\nu,
\label{eqn:e_recombination}
\end{equation}

\noindent where the reaction rate coefficient from $j+1$ to $j$ is labeled $k_e^{j}$   Ions can charge exchange with atomic hydrogen,

\begin{equation}
C^{(j+1)+} + H \rightarrow C^{j+} +H^+.
\label{eqn:CX_H}
\end{equation}

\noindent where the reaction rate coefficient from $j+1$ to $j$ is labeled $k_{cx}^{j}$.  In regions with molecular hydrogen C$^{j+}$ generally reacts rapidly via ion-molecule reactions, 

\begin{eqnarray}
&C^{(j+1)+} + H_2 &\rightarrow C^{j+} +H_2^+\\
&C^{(j+1)+} + H_2  &\rightarrow C^{(j-1)+} +H^+ + H^+\\
&C^{(j+1)+} + H_2 &\rightarrow C^{j+} +H^+ + H
\label{eqn:C_H2}
\end{eqnarray}

\noindent except for C$^+$ which reacts slowly with H$_2$ via radiative association, 
\begin{equation}
C^+ + H_2 \rightarrow CH_{2}^{+} + h\nu,
\label{eqn:C+_H2}
\end{equation}  

\noindent we label these ion molecule reactions $k_{im}^j$. 
To calculate the number density of carbon ions we need to solve a set of balance equations,

\begin{eqnarray}
(\zeta_{XL}^{j} + \zeta_{XK}^{j}+\beta^{j-1}) n_{j}(C^{j+}) = \zeta_{XK}^{j-2} n_j(C^{(j-2)}) \nonumber \\ 
+\zeta_{XL}^{j-1} n_j(C^{(j-1)+}) +  \beta^{j} n_{j+1}(C^{(j+1)+})
\label{eqn:balance_equations}
\end{eqnarray}

\noindent where

\begin{equation}
\beta^j =  k_{e}^{j} n(\rm {e}) + k_{CX}^{j} n(\rm {H}) + k_{im}^{j} n(\rm{H_2}).
\end{equation}

\noindent for $j$ = 0 to $j_{max}$-1 (note that $\zeta^{j_{max}}$ = 0  and $\beta^{j_{max}}$= 0 in Equation~\ref{eqn:balance_equations} because there is no ionization out of and no recombination into the maximum charge state, respectively, and that $\beta^{-1}$ = 0 because there is no recombination from the lowest charge state).  In addition we need the number density conservation equation,

\begin{equation}
\sum\limits_{0}^{j_{max}} n_j(C^{j+}) = n_t(C)
\end{equation}

\noindent where $n_t$(C) is the total number density of all carbon charge states.  For the galactic center interstellar medium the X-ray energy range of interest  is greater than 1 keV and we can neglect the L-shell photoionization rate (see Section~\ref{sec:ionization_rates}).  These equations can be solved analytically and used to calculate the fractional ionization of each species, $f_i$ = $n_i/n_t$, where $i$ labels the ionization state, and $\sum\limits_{0}^{j_{max}}  f_i$ = 1.

\subsection{X-ray photoionization cross sections}
\label{sec:ionization_cross}

\cite{Verner1993} provide fits for X-ray photoionization cross sections for C$^0$ to C$^{4+}$ for the 1s-shell (K-shell), C$^0$ to C$^{+3}$ for the 2s-shell (L-shell), and  C$^0$ and C$^+$ for the 2-p shell (L-shell). The 2s- and 2p-shell cross sections contribute less than 5\% to the total cross section above 0.4 keV.  To facilitate solving for $f_i$ we fit the cross sections for X-ray photoionization of C$^0$, C$^+$, C$^{2+}$, C$^{3+}$, and C$^{4+}$ from \cite{Verner1993} and \cite{Verner1995} with a power law fit above 0.4 keV.  The fits are very similar for all the K-shell reactions C$^{j+}$ $\rightarrow$ C$^{(j+2)+}$, and have a form, $\sigma(E) = \sigma_0 E_{keV}^{-b}$, and for simplicity we adopt an average value, $\sigma_0 = 4.1\times 10^{-20}$ cm$^2$ and  $b =$ 2.8.  In Section~\ref{sec:results} we will combine $\sigma(E)$ with the X-ray photon flux to calculate the X-ray photoionization rates. As we are only interested in solving for the effects of X-rays on the carbon ionization balance in regions where neutral carbon is readily ionized by UV we will neglect the balance equations for C$^0$.
This assumption is valid in the presence of UV which ionizes neutral carbon rapidly in unshielded regions in the interstellar medium, otherwise there would not be any C$^+$ and it would recombine to C$^0$. Furthermore, the ionization of C$^0$ by UV is generally orders of magnitude larger than that by X-rays.

\subsection{Electron recombination reaction rate coefficients}

The reaction rate coefficients for radiative and dielectronic recombination of electrons with carbon ions up to C$^{6+}$ have been calculated by \cite{Nahar1997} and \cite{Badnell2003} and \cite{Badnell2006} with radiative recombination dominating at low temperatures (as in diffuse clouds) but dielectronic recombination becoming more important at the higher temperatures found in the WIM and DIS.  We use the values for the total recombination rate coefficients published in Table 5 of \cite{Nahar1997} in our model calculations. The reactions are given in Table~\ref{tab:Table1} and we list the reaction rate coefficients for two characteristic gas temperatures, 100 K and 8000 K.  

\begin{table}[htbp]																	
\caption{Electron recombination reactions}
\label{tab:Table1}															
		\begin{tabular}{lccc}																	
		\hline	
Reaction$^a$ & 100K  &  8000K \\
 \hline
\hline
C$^+$ + e $\rightarrow$ C$^0$ + h$\nu$  & 9.1$\times$10$^{-12}$ & 7.3$\times$10$^{-13}$ \\
C$^{2+}$ + e $\rightarrow$ C$^+$ + h$\nu$  & 4.2$\times$10$^{-11}$ & 6.9$\times$10$^{-12}$  \\
C$^{3+}$ + e $\rightarrow$ C$^{2+}$ + h$\nu$  & 3.2$\times$10$^{-10}$ & 1.8$\times$10$^{-11}$ \\
C$^{4+}$ + e $\rightarrow$ C$^{3+}$ + h$\nu$  & 1.6$\times$10$^{-10}$ & 1.0$\times$10$^{-11}$ \\
\hline
\end{tabular}
\\	
a. The reaction rate coefficients, in cm$^3$ s$^{-1}$, are from Table 5 of \cite{Nahar1997}. 
 \end{table}

\subsection{Charge exchange with atomic hydrogen}
Reaction rate coefficients for charge transfer between carbon ions and atomic hydrogen have been summarized  for C$^+$ to C$^{4+}$ by \cite{Kingdon1996} along with fits in the form, 

\begin{equation}
k_{CX}(T_4) = aT_{4}^{b}[1 + c \rm{e}^{dT_4}]\,\, \rm{cm^3 s^{-1}}
\end{equation}

\noindent where $T_4$ is the kinetic temperature in units of 10$^4$ K, and the fitting coefficients are given in their Table 1.  The charge exchange of C$^+$ with H is very small owing to the large endothermic barrier, $\Delta$E$\sim$ 2.5 eV (due to the different ionization potentials of C and H).  Typically most exothermic charge exchange reactions of highly ionized metals proceed rapidly under ISM conditions with a reaction rate coefficient of order 10$^{-9}$ cm$^3$ s$^{-1}$, but that is not the case for doubly ionized carbon. Although charge exchange of C$^{2+}$ with H is energetically favorable, its cross section is small owing to the nature of the crossings in the intermediate molecular potential \citep{McCarroll1975,Kingdon1996}. However, the higher ionization states of carbon do charge exchange rapidly.  In Table~\ref{tab:Table2} we list the reaction rate coefficients for carbon ions at 8000 K and 100 K, typical of temperatures found in the warm ionized medium and diffuse atomic hydrogen gas, respectively.

\begin{table}[htbp]																	
\caption{Charge exchange with H reaction rate coefficients}
\label{tab:Table2}															
		\begin{tabular}{lccc}																	
\hline	
Reaction &  $k_{CX}$(8000 K)$^a$  &  $k_{CX}$(100 K) $^a$&  \\
  \hline
\hline
C$^+$ + H $\rightarrow$ C$^0$ + H$^+$  & $\sim$7$\times$10$^{-18}$  & $<<$10$^{-30}$ \\
C$^{2+}$ + H $\rightarrow$ C$^+$ + H$^+$  & 1.1$\times$10$^{-12}$  & 1.3$\times$10$^{-16}$  \\
C$^{3+}$ + H $\rightarrow$ C$^{2+}$ + H$^+$  & 3.1$\times$10$^{-9}$  & 1.5$\times$10$^{-9}$   \\
C$^{4+}$ + H $\rightarrow$ C$^{3+}$ + H$^+$  & 2.1$\times$10$^{-9}$  & 2.8$\times$10$^{-9}$   \\
\hline
\end{tabular}
\\	
a.  In units of cm$^3$ s$^{-1}$.
\end{table}

\subsection{C$^+$ ion molecule reactions with H$_2$}

C$^+$ reacts slowly with molecular hydrogen via associative recombination, C$^+$ + H$_2 \rightarrow$ CH$_{2}^{+}$ + h$\nu$ \citep{McElroy2013}, and this reaction does not play a significant role in the carbon ionization balance outside of dense PDRs.  In contrast, the reactions of C$^{n+}$ with H$_2 $ for $n \ge 2$ are likely to be fast, $\sim$10$^{-10}$ to 10$^{-9}$ cm$^3$ s$^{-1}$ for most cases \citep{Langer1978}.  Thus the admixture of a small amount of molecular hydrogen can have a large effect on the carbon X-ray photoionization balance.  The only multiply charged carbon ion molecule reaction that has been measured in the lab is, C$^{2+}$ + H$_2$ \citep{Smith1981} and it has a reaction rate coefficient of $\sim$1.2$\times$10$^{-10}$ cm$^3$ s$^{-1}$ with the dominant channel $\rightarrow$ H$^+$ + H$^+$ + C comprising  $\sim$90\% of the total. We assume that higher charge states have the same reaction rate coefficient and end products, although the actual product channel is not critical for the model, and these are listed in Table~\ref{tab:Table3}.

\begin{table}[htbp]																	
\caption{Reaction rate coefficients of C$^{n+}$ with H$_2$}
\label{tab:Table3}															
		\begin{tabular}{lcc}																	
\hline	
Reaction &  $k$(100K)$^a$  &   \\
  \hline
\hline
C$^+$ + H$_2$ $\rightarrow$ CH$_{2}^{+}$ + h$\nu$   & 6$\times$10$^{-16}$  \\
C$^{2+}$ + H$_2$ $\rightarrow$ C$^{0}$ + H$^+$ + H$^+$& 1.2$\times$10$^{-10}$     \\
C$^{3+}$ + H$_2$ $\rightarrow$ C$^{+}$ + H$^+$ + H$^+$  & 1.2$\times$10$^{-10}$     \\
C$^{4+}$ + H$_2$ $\rightarrow$ C$^{2+}$ + H$^+$ + H$^+$  & 1.2$\times$10$^{-10}$      \\
\hline
\end{tabular}
\\	
a.  In units of cm$^3$ s$^{-1}$.
\end{table}



\section{Results}
\label{sec:results}

The gas in galactic nuclei is exposed to a different radiation environment than that in the galactic disks with intense  UV and X-ray radiation fields due to the presence of  young O-type stars, supernovae, stellar black holes, and a massive black hole at the center.  In AGNs this environment reaches an extreme with X-ray luminosities, $L$(X-ray),  in excess of 10$^{40}$ erg s$^{-1}$ and up to $\sim$2$\times$10$^{47}$ erg s$^{-1}$ \citep[e.g.][]{Ebrero2009}. These large luminosities are sufficient to influence the distribution of higher ionized states of carbon in galactic nuclei and depends on the flux of soft X-rays with energies above the K-shell ionization threshold ($\sim$0.3 - 0.4 keV). 
If the source of X-rays in the galactic center is due to an accreting black hole the X-ray luminosity flux, 

\begin{equation}
F_X=\frac{L(\mbox{X-ray})}{r^2}.
\end{equation}

\noindent In the notation of \cite{Maloney1996},  $F_X \sim$ 8.4$\times$10$^5 L_{44}r_{\rm pc}^{-2}$ erg cm$^{-2}$ s$^{-1}$, where $L_{44}$ = $L_X/10^{44}$ erg s$^{-1}$ and $r_{\rm pc}$ is the distance to the X-ray source in pc.  Thus, neglecting absorption above 1 keV, $F_X$ can be very large throughout most of a galactic nucleus and these X-rays will have a major impact on the properties of the neutral and ionized gas. While X-ray emission from massive black holes can effect the ionization balance, chemistry, and thermal properties over large volumes of the galactic center, other weaker sources can have comparable influence on nearby clouds.  For example, strong supernova shocks can produce luminosities $L_X \sim$ 10$^{36}$ to 10$^{40}$ erg s$^{-1}$ and stellar black holes can produce $\sim$10$^{37}$ -- 10$^{38}$ erg s$^{-1}$  \citep{Heindl1993}.

\cite{Maloney1996}  modeled the effects of enhanced UV and  X-ray fluxes on the thermal and chemical balance of molecular gas in galactic nuclei, generating a grid of XDR 
models that covered a range of X-ray radiation field energy fluxes $F_X$ = 0.1 to 10$^6$ erg cm$^{-2}$ s$^{-1}$.  They set a lower limit of 1 keV to guarantee that the X-rays will penetrate sufficient depth into the molecular cloud to affect the chemical and thermal balance. \cite{Meijerink2007} also modeled the effects of enhanced UV and  X-rays on molecular gas in galactic nuclei, but covered a narrower range of energy flux, $F_X$ = 1.6$\times$10$^{-2}$ to 1.6$\times$10$^2$ erg cm$^{-2}$ s$^{-1}$.   Here we consider the effects of large X-ray fluxes on the carbon balance in the ionized ISM, the diffuse atomic hydrogen clouds, and the photodissociation regions (PDRs) of dense molecular clouds in galactic nuclei. We begin by calculating the X-ray photoionization rates.
  
\subsection{X-ray photoionization rates}
\label{sec:ionization_rates}

Here we calculate the X-ray ionization rates for carbon ions parameterized in terms of the X-ray energy flux, $F_X$. The X-ray photoionization rate for carbon ions of charge state $j$ is

\begin{equation}
\zeta^j(C^{j+}) =\int_{E_{min}}^{E_{max}} \sigma_j(E)\frac{dJ}{dE}dE,
\label{eqn:ionization}
\end{equation}

\noindent where $dJ/dE$ is the photon spectral distribution of X-rays  as a function of energy  in units of photons cm$^{-2}$ s$^{-1}$ keV$^{-1}$, and $E_{min}$ and $E_{max}$ are the minimum and maximum X-ray energies, respectively.   The spectral distribution is usually expressed as a power law distribution, $dJ/dE =  J_0/E_{{\rm keV}}^{\Gamma}$.   For Seyfert galaxies $\Gamma$ is typically in the range 1.4 to 2.0 below 100 keV with a distribution average $\sim$1.7 \citep{Mushotzky1993}.  However, the spectral index can vary with energy, for example \cite{Iwasawa2011} found different spectral slopes for hard and soft X-rays in a sample of LIRGs.  For active galactic nuclei  \citep{Maloney1996} adopted $\Gamma \sim$ 2 for AGNs and \cite{Meijerink2007}  $\Gamma =$ 1.9 for the X-ray spectrum of an accreting massive black hole. Here we will adopt  $\Gamma$ = 1.9.  We will set $E_{min}$ = 1 keV because, as discussed above, lower energy X-rays can be neglected except very near the X-ray source, and $E_{max}$ is usually taken to be 100 keV.


To calculate the photoionization rate as a function of the energy flux we will vary $J_0$ to represent different X-ray fluxes. We rewrite the spectral energy distribution as,

\begin{equation}
\frac{dJ}{dE} = \frac{F_X J_0(1)}{E_{\rm keV}^\Gamma}
\label{eqn:dJdE}
\end{equation}

\noindent where $J_0$(1) is the value that produces a luminosity flux of 1 erg cm$^{-2}$ s$^{-1}$ for a given spectral index $\Gamma$. We find $J_0$(1) by solving

\begin{equation}
\int_{E_{min}}^{E_{max}} E\frac{dJ}{dE}dE = \int_{E_{min}}^{E_{max}} E\frac{J_0(1)}{E_{keV}^\Gamma}dE = 1\,\, {\rm (erg\, cm^{-2}\,\, s^{-1})},
\end{equation}

\noindent substituting $\Gamma$ =1.9,  integrating from 1 to 100 keV, and converting units from keV to ergs.  This conversion yields, $J_0$(1) = 1.1$\times$10$^8$ photons cm$^{-2}$ s   $^{-1}$ keV$^{-1}$.  We now express $dJ/dE$ = 1.1$\times$10$^8$$F_X/E^{1.9}$ where $F_X$ is in units of erg cm$^{-2}$ s$^{-1}$.  Substituting $dJ/dE$  and  $\sigma (E_{keV})$  into Equation~\ref{eqn:ionization} yields

\begin{equation}
\zeta^j(C^j) = 1.1\times 10^8 F_X \sigma_0 \int_{1}^{100} \frac{dE}{E^{\Gamma+ b}}.
\label{eqn:ionization_final}
\end{equation}

\noindent This equation is valid at the boundary of the cloud as long as we can neglect absorption of the X-rays.  For 1 keV, our assumed lower threshold, where most of the ionization occurs, the absorption opacity is less than one for a hydrogen column density $\lesssim$ (5 - 10)$\times$10$^{21}$ cm$^{-2}$, depending on the metallicity. Only the GMCs have sufficient column density to absorb X-rays at this energy, however, as discussed the X-ray ionization has little effect on the \cii emission from the molecular hydrogen boundaries of the cloud and therefore we neglect the attenuation of X-rays in determining the luminosity from this layer. Substituting $\sigma_0$, $\Gamma$, and $b$ into Equation~\ref{eqn:ionization_final}  and integrating from 1 to 100 keV yields, $\zeta^j(C^j)$ $\simeq$ 1.2$\times$10$^{-12}$ $F_X$ s$^{-1}$. 

In the neutral gas clouds the electrons ejected by X-ray photoionization primarily of hydrogen produce secondary ionization. The multiplicative effects of these secondary ionizations of metals is of order a few and here we adopt a factor of two for carbon \citep[see][]{Maloney1996,Meijerink2005}. However, in the ionized WIM and DIS the photoejected electrons scatter via Coulomb interactions off the ambient electrons.  This scattering is large compared with the electron ionization cross sections for C$^{n+}$ \cite[cf.][]{Suno2006} and the photoejected electrons  are  rapidly thermalized. Therefore above a fractional ionization of a couple of percent, thermalization  is so rapid \citep{Maloney1996} that the probability of secondary ionization of carbon can be neglected in the WIM and DIS. 

The X-ray photoionization rate of carbon ions can be quite large in galactic nuclei and this process will result in the conversion of C$^+$ to higher ionization states in diffuse ionized regions.  \cite{Meijerink2007} considered $F_X$ = 1.6$\times$10$^{-2}$ to 1.6$\times$10$^2$ erg cm$^{-2}$ s$^{-1}$ in their XDR models of dense gas in galaxy nuclei and \cite{Maloney1996} studied a range of 0.1 to 10$^6$ erg cm$^{-2}$ s$^{-1}$.  In this Section, for illustrative purposes, we will usually consider an X-ray flux of $F_X$ = 10$^{-3}$ to 10$^4$  erg cm$^{-2}$ s$^{-1}$, but sometimes consider higher values. The lower limit reflects an ionization rate at which X-rays are unimportant in the ion balance. 
In the following subsections we present the distribution of ionized states of carbon as a function of X-ray flux, under different conditions of electron, atomic hydrogen, and molecular hydrogen abundances, corresponding to different AGN ISM environments. To simplify the solution we  include an ionization state only up to C$^{4+}$, as our main purpose is to study the conversion of C$^+$ to higher states.  Thus the fractional abundance $f$(C$^{4+}$) represents all charge states $z \ge$4.

\subsection{Highly ionized regions}
\label{sec:Dense_WIM}

There are several  environments in which hydrogen is completely ionized.  One is the warm ionized medium, where the densities in the disk are usually low, $n$(e) $\sim$ few$\times$10$^{-2}$ cm$^{-3}$ \citep{Haffner2009}, but are much higher in galactic nuclei. In the Central Molecular Zone, $\sim$ 10 cm$^{-3}$ \citep{Cordes2003,Roy2013}, but may be higher in AGNs.  Other highly ionized regions are those associated with \hii sources and \hiino-like dense ionized skins (DIS) around dense molecular clouds, where the electron densities may be high, $n$(e) $\sim$ 1 - 100 cm$^{-3}$ \citep{Oberst2011,Langer2015}. Finally, there is the hot ionized medium (HIM) which is detected via thermal X-ray emission and where hydrogen and helium are completely ionized.  The HIM would have $T_{\rm kin} \sim$ 10$^5$ - few $\times$10$^6$ K and $n$(e) $<$ few$\times$10$^{-3}$ cm$^{-3}$ \citep{Cox2005,Ferriere2007}. In general the densities  for the HIM are too small for this component to contribute much to the \cii emission from galactic nuclei and we neglect  it here.  

\begin{figure}
 \centering
      \includegraphics[width=6.70cm]{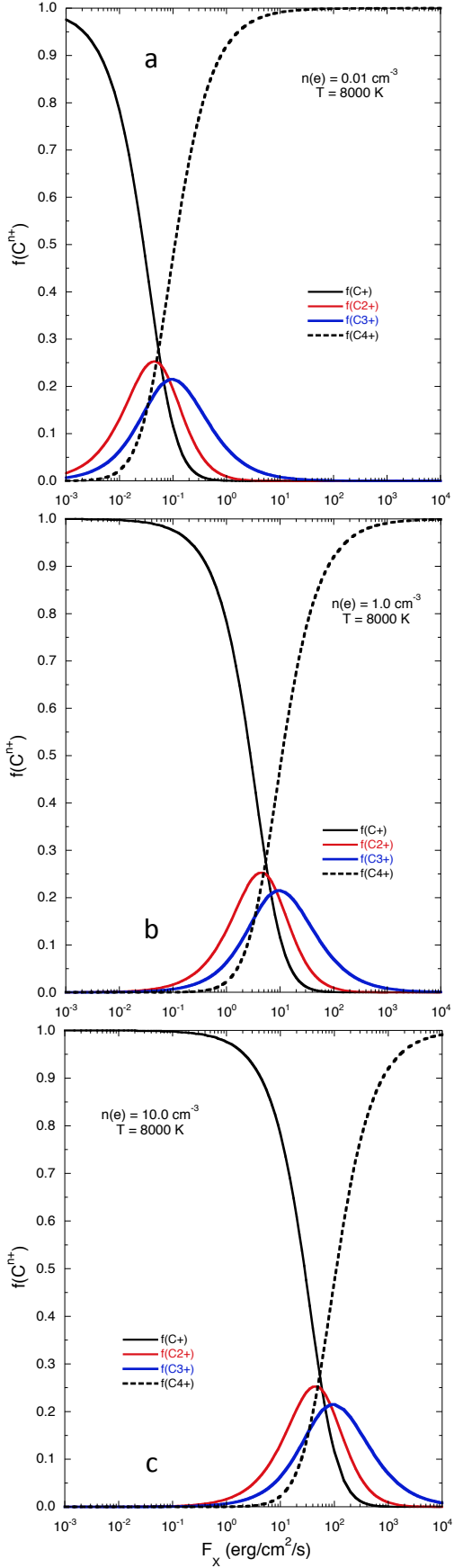}
      \caption{Fractional ionization state of C$^+$ to C$^{4+}$ versus X-ray luminosity flux, $F_X$, for  $n$(e)= 0.01 cm$^{-3}$ (top), 1.0 cm$^{-3}$ (middle), and 10.0 cm$^{-3}$ (bottom) at $T_{\rm kin}$ = 8000 K, assuming a fully ionized region ($n$(H) = $n$(H$_2$) = 0 cm$^{-3}$).}
                           \label{fig:fig2}
 \end{figure}

In Fig.~\ref{fig:fig2}(a) we show the solutions for  a low density completely ionized gas, $n$(e) = 0.01 cm$^{-3}$ (fully ionized, $n$(H) = $n$(H$_2$) = 0 cm$^{-3}$) and $T_{\rm kin}$ = 8000 K. It can be seen that as the X-ray energy flux increases the C$^+$ is first converted to C$^{2+}$, then C$^{3+}$, and at $F_X$ $\sim$10$^{-1}$ erg cm$^{-2}$ s$^{-1}$ C$^+$ becomes negligible and all the carbon is in the higher ionization states.  For $F_X >$ 1 erg cm$^{-2}$ s$^{-1}$ carbon is essentially in the highest charge state considered here, C$^{4+}$. For galactic centers and \hiino-like regions where the electron densities are higher,  we show results for $n$(e) = 1 cm$^{-3}$ and 10 cm$^{-3}$ in Figs.~\ref{fig:fig2}(b)  and (c), respectively. Increasing the electron density shifts the transition from C$^+$ to higher ionization states to larger $F_X$ as a larger X-ray radiation flux is needed to offset the more rapid electron recombination to lower ionization states.  For all the electron densities considered here there is a threshold in $F_X$ where C$^+$ goes to zero in the highly ionized gas.  Adding a small amount of atomic hydrogen, $n$(H)$\sim$ 0.01 cm$^{-3}$, does not have much of an effect on the distribution of ionization states of carbon (not shown).  For the hot ionized medium the combination of a lower electron recombination rate coefficient at the higher temperatures and the lower electron density pushes the transition from C$^+$ to C$^{j+}$ to low values of $F_X$; for example, for $T_{\rm kin}$ = 10$^5$ K and $n$(e) = 0.003 cm$^{-3}$ we find $f$(C$^+$) $<$ 0.1 for the lowest flux considered here $F_X$ = 10$^{-3}$ erg cm$^{-2}$ s$^{-1}$.


\subsection{Atomic hydrogen clouds}
\label{sec:atomic_H_clouds}

\begin{figure}
 \centering
            \includegraphics[width=6.7cm]{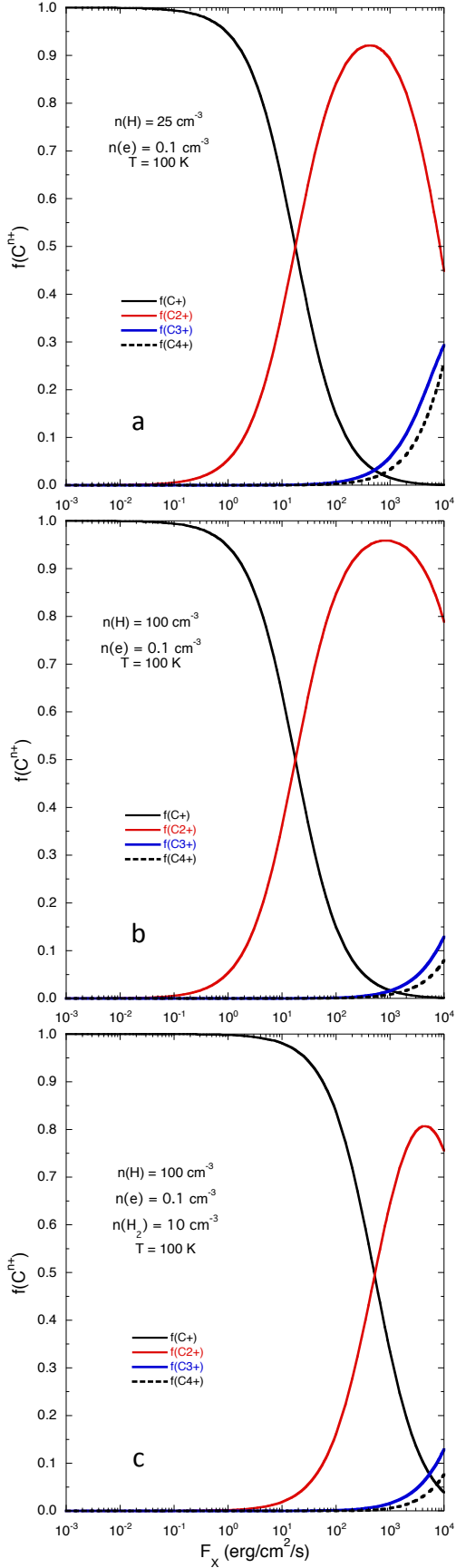}
      \caption{Fractional ionization state  of C$^+$ through C$^{4+}$ versus X-ray luminosity flux, $F_X$, for a diffuse atomic hydrogen cloud with $n$(H) = 25 cm$^{-3}$ (panel a) and 100 cm$^{-3}$ (panel b), assuming $n$(e) = 0.1 cm$^{-3}$, and $T_{\rm kin}$ = 100 K and no H$_2$, panel (c) includes a 10\% admixture of H$_2$.}
      \label{fig:fig3}
 \end{figure}

We next address the question whether a large flux of X-rays will modify the distribution of C$^+$ in low density diffuse atomic hydrogen clouds.  The presence of even modest amounts of atomic hydrogen will keep carbon from reaching highly ionized states because of the efficiency of charge exchange, except for C$^{2+}$ which does not charge exchange efficiently.  In Fig.~\ref{fig:fig3}(a)  we show the fraction of C$^+$ and higher ionization states for conditions in a diffuse atomic hydrogen cloud with $T_{\rm kin}$ =100 K, a hydrogen density $n$(H) = 25 cm$^{-3}$, a small admixture of electrons, and no molecular hydrogen.  As can be seen in this figure only the two lowest carbon ionization states are present, and the higher charge states have very small fractional abundances because they charge exchange rapidly with H to lower ionization states.  For $F_X \lesssim$ 10 erg cm$^{-2}$ s$^{-1}$ carbon is mainly singly ionized, however in the presence of an X-ray energy flux $F_X \gtrsim$ 10 erg cm$^{-2}$ s$^{-1}$ X-ray K-shell photoionization converts C$^+$ rapidly into C$^{3+}$, which then charge exchanges to C$^{2+}$. 
Thus under high X-ray flux conditions low density atomic hydrogen clouds would have  a low fraction of C$^+$ but a high fraction of C$^{2+}$  and \cii emission would be reduced. In Fig.~\ref{fig:fig3}(b)  we show the same calculation for a hydrogen density $n$(H) =100 cm$^{-3}$ and we see it has the same behavior as the lower $n$(H) case but the transition occurs at a higher flux $F_X$.  

The results in  Fig.~\ref{fig:fig3} assume a very small, but non-neglible, electron abundance because it will be present in atomic clouds due to ionization of hydrogen by cosmic rays and X-rays.  Under typical ISM conditions found in the Galactic disk the electron fraction in diffuse atomic clouds is small ($n$(e)/$n$(H)$ <$10$^{-3}$).  However, the electron fraction increases with increasing X-ray flux. We can estimate the electron fraction by balancing X-ray ionization with electron recombination.  We find  a fractional electron abundance $\sim$10$^{-2}$ for $F_X$ = 10$^3$ erg cm$^{-2}$ s$^{-1}$ in a cloud with $n$(H) = 10$^2$ cm$^{-3}$.  However, the addition of a small fractional abundance of electrons ($<$ 0.1) does not change the distribution of C$^{n+}$ in the diffuse clouds.  Finally, we note that for very large X-ray energy fluxes, $F_X >$ 10$^5$ erg cm$^{-2}$ s$^{-1}$ and for $n$(H) $<$ 100 cm$^{-3}$  that the fractional abundance of H$^+$, $f$(H$^+$) $>$ 0.1 and the assumption of a neutral atomic hydrogen cloud is no longer valid.


\subsection{Molecular hydrogen clouds}

Highly ionized states of carbon react rapidly with molecular hydrogen and so the main reservoirs of carbon in molecular cloud PDR and XDR regions is C$^+$.  \cite{Meijerink2005} and \cite{Meijerink2007} included the production of C$^{2+}$ by X-ray photoionization in their XDR models (see \cite{Meijerink2005} Appendix D.3.1).  However, as discussed above, this rapid reaction reduces the C$^{2+}$ abundance to very low values in their models, even at their maximum energy flux, $F_X$ = 160 erg cm$^{-2}$ s$^{-1}$. We illustrate the influence of a small admixture of molecular hydrogen, $n$(H$_2$) = 10 cm$^{-3}$, or a 10\% admixture,  on the carbon ionization balance in a diffuse hydrogen cloud with $n$(H) = 100 cm$^{-3}$ and $T_{\rm kin}$ = 100K.  As seen in Fig.~\ref{fig:fig3}(c) the crossover from C$^+$ to C$^{2+}$ occurs at a higher value of  $F_X$ than in a cloud with only atomic hydrogen (Fig.~\ref{fig:fig3}(b)). 

\begin{figure}
 \centering
      \includegraphics[width=8cm]{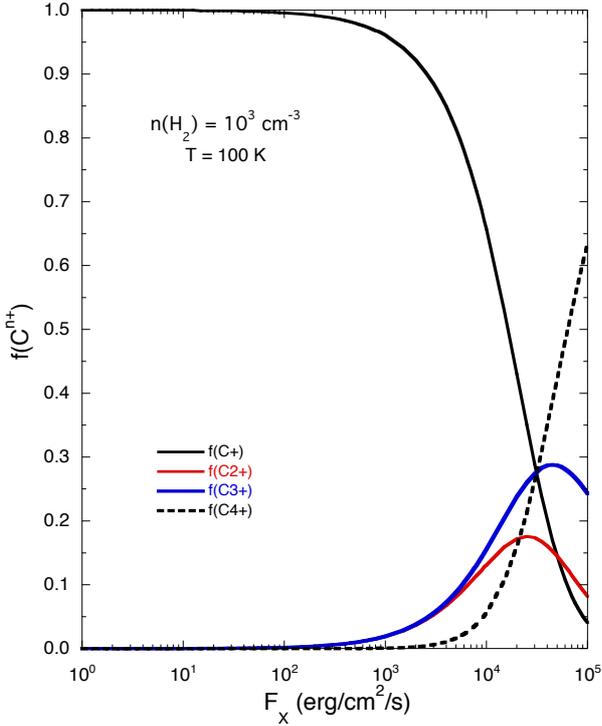}
      \caption{Fraction of  carbon ionization states versus  X-ray luminosity flux, $F_X$, for a dense molecular hydrogen gas, $n$(H$_2$) = 10$^3$ cm$^{-3}$ and $T_{\rm kin}$ =100 K, with an admixture $n$(H) = 1 cm$^{-3}$ and $n$(e) = 0.1 cm$^{-3}$.  Carbon remains in the singly ionized state, C$^+$, up to $F_X \sim$ 5$\times$10$^3$ erg cm$^{-2}$ s$^{-1}$. }
       \label{fig:fig4}
 \end{figure}

The rapid charge transfer with molecular hydrogen has a more profound effect in dense PDRs where the H$_2$ suppresses significantly the population of higher carbon ionization states as is illustrated in Fig.~\ref{fig:fig4} for n(H$_2$) = 10$^3$ cm$^{-3}$, where the hydrogen density is held constant (which may not be valid at high X-ray fluxes as discussed below).  In the XDR models of \cite{Maloney1996,Maloney1997} and \cite{Meijerink2005,Meijerink2007}  the molecular hydrogen density $n$(H$_2$) $\ge$ 10$^3$ cm$^{-3}$ and  molecular hydrogen is efficient at suppressing the highly ionized states of carbon to such an extent that is necessary to go to very large X-ray fluxes, $F_X > $10$^3$ erg cm$^{-2}$ s$^{-1}$  to see any significant decrease in C$^+$.  The results are relatively insensitive to the choice of kinetic temperature because the charge exchange reactions with H$_2$ dominate the ionization balance in the PDRs .

In Fig.~\ref{fig:fig4} the hydrogen density was held constant to illustrate the effects of the reactions of the multiply ionized carbon with the ambient gas. However, the X-rays not only ionize carbon but also ionize the hydrogen molecules, producing $H_2^+$ which can subsequently be destroyed in reactions with H$_2$ leading to atomic hydrogen (H$_2^+$ + H$_2 \rightarrow$ H$_3^+$ + H, followed by reactions of H$_3^+$ with atoms, trace molecules and electrons that lead to additional production of H atoms). \cite{Maloney1996} estimate that every direct ionization of H$_2$ leads to three H atoms.  The formation of H$_2$ takes place on grain surfaces and the formation rate depends on the grain cross section, sticking coefficients, and recombination efficiency as summarized in Appendix D.1 of \cite{Meijerink2005}.  Typical formation rates are of order (1 - 3)$\times$10$^{-17}$$n$(H)$n_{tot}$, where $n_{tot}$ is the total number of hydrogen nucleons (= $n$(H) + 2$n$(H$_2$)). Thus the PDR layers of dense molecular clouds become atomic hydrogen (and eventually H$^+$) above a threshold $F_X$.  We estimate that for dense PDR layers the fraction of H, $f$(H) is of order one for  $F_X$ $\gtrsim$ 10$^2$  erg cm$^{-2}$ s$^{-1}$. Detailed models of PDR/XDR layers show this transition \citep[see][]{Maloney1996,Meijerink2005}.  Thus, although X-rays can, in principle, produce multiply ionized carbon ions in the H$_2$ gas, under realistic conditions in dense molecular clouds this situation is unlikely to be maintained as the molecular hydrogen is dissociated at a lower energy flux than needed to sustain highly ionized states of carbon.  As discussed below this process results in the PDR/XDR layers occurring deeper in the cloud where the X-ray flux has been attenuated.




 
\subsection{X-ray ionization of nitrogen and oxygen}

In addition to carbon, atomic oxygen, and nitrogen and oxygen  ions have fine-structure transitions that are important diagnostics of the ISM in galaxies and are used as tracers of the star formation rate \cite[see, for example][]{DeLooze2014}. These atoms and ions also play an important role in the thermal balance of the galactic gas. The O and N atomic and ionic lines also show intensity deficits  \citep[e.g.][]{Luhman1998,Dale2004,Gracia-Carpio2011}. 


In this paper we do not model the complete ionization distribution of these species and their far-infrared emission but instead focus on a few key differences in their chemistry in highly ionized gas as it is likely that much of the emission from galaxies come from these ISM components.  These differences alter the distribution of their higher ionization states as compared with carbon\footnote{An example of the differences in the distribution of C, N, and O ionization states in the presence of X-rays can be seen in the models of \cite{Adamkovics2011} for protoplanetary disks.}.  First, the ionization potential of  N (14.534 eV) is significantly greater than hydrogen (13.598 eV) so  photoionization occurs only for EUV photons somewhat beyond the Lyman limit (911.8 \AA ngstr\"{o}m).  In contrast the ionization potential of oxygen, 13.6181 eV, is only slightly above that of hydrogen but still beyond the Lyman limit.  In the warm ionized medium charge exchange reactions with protons are able to sustain a significant fraction of O$^+$ and N$^+$ even in the absence of EUV or X-rays.  In the case of oxygen the energy difference is small (in temperature units $\sim$ 224K) that the protons in the warm and hot ISM can efficiently charge exchange with  O, but in the case of nitrogen the energy difference is so large ($\sim$10,980 K) that the ionized gas must be $\gtrsim$ few$\times$10$^3$ K for efficient ionization of N. Furthermore, calculations of the cross sections for H$^+$ + N $\rightarrow$ N$^+$ + H differ \citep{Kingdon1996,Lin2005} resulting in somewhat different reaction rates at the temperatures of interest in the WIM \citep{Langer2015}.  

In ISM gas with atomic hydrogen, the ions will neutralized by charge exchange with H because the reaction rate coefficient for O$^+$ + H is fast, $\sim$10$^{-9}$ cm$^{3}$ s$^{-1}$, and, while  that for N$^+$ + H is slow, $\sim$10$^{-12}$ cm$^{3}$ s$^{-1}$ at 8000 K, it is still faster than that for C$^+$. Furthermore, in contrast to C$^{2+}$, the charge exchange cross sections for N$^{2+}$ and O$^{2+}$ with H are fast, $\sim$10$^{-9}$ cm$^{3}$ s$^{-1}$ at 8000 K, so higher ionization states of N and O will be difficult to sustain in gas containing atomic hydrogen. It will be even more difficult to have an abundance of N$^+$ and O$^+$ in molecular hydrogen gas as their reaction rate coefficients with H$_2$ are very large and, indeed, are primary pathways to forming oxygen- and nitrogen-bearing molecules in shielded regions. Therefore, highly ionized states of nitrogen and oxygen are likely to be found only in the warm highly ionized gas.

\begin{figure}
 \centering
      \includegraphics[width=6.7cm]{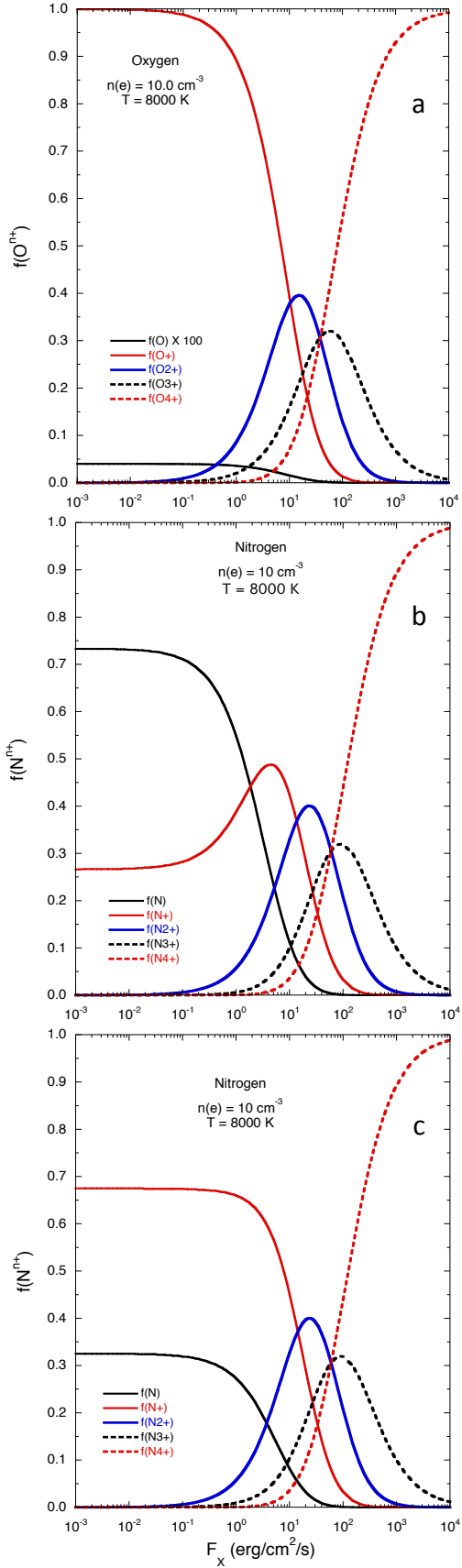}
      \caption{Fraction of  oxygen ionization states (a) and nitrogen ionization states (b and c) versus X-ray luminosity flux, $F_X$, for a dense WIM with $n$(e) = 10 cm$^{-3}$ and T$_k$=8000 K. Panel (b) assumes the charge transfer rates of \cite{Lin2005} and panel (c) those of \cite{Kingdon1996} - see text. }
       \label{fig:fig5}
 \end{figure}

We have modeled the distribution of the ionized states of nitrogen and oxygen under conditions of a warm dense ionized medium, similar to the model for carbon in Section~\ref{sec:Dense_WIM}.  We use the photoionization cross-sections in \cite{Verner1993}, charge exchange reaction rate coefficients from \cite{McElroy2013}, \cite{Lin2005}, and \cite{Kingdon1996}, and electron recombination rate coefficients from \cite{Nahar1997} and \cite{Badnell2003,Badnell2006}.  We show in Fig.~\ref{fig:fig5} results for nitrogen and oxygen in a dense ionized gas with $n$(e) = 10 cm$^{-3}$ and $n$(H$^+$) = $n$(e).  Fig.~\ref{fig:fig5}(a)  is a plot of the fractional abundances of oxygen ions, O$^0$ to O$^{4+}$.  It can be seen that proton charge exchange keeps oxygen ionized even in the absence of X-rays and that neutral oxygen is essentially absent. As the X-ray flux increases  O$^{2+}$ increases at the expense of O$^+$ and for $F_X >$ few$\times$10$^2$ erg cm$^{-2}$ s$^{-1}$ all the oxygen is in the highest ionization state considered here, O$^{4+}$.  Fig.~\ref{fig:fig5}(b) and (c) plot the distribution of nitrogen ions, N$^0$ to N$^{4+}$, using the charge exchange rate coefficients of \cite{Lin2005} and \cite{Kingdon1996}, respectively. Their distribution is somewhat different than oxygen at low X-ray fluxes because the temperature is not high enough for efficient charge exchange by H$^+$ to keep all the nitrogen ionized.  Instead the nitrogen is only partially ionized, but more so using the results of \cite{Kingdon1996}.  As the X-ray flux increases it ionizes the nitrogen and N$^+$ increases, and then at larger fluxes the distribution starts to resemble those of carbon and oxygen.  For $F_X >$ few$\times$10$^2$ erg cm$^{-2}$ s$^{-1}$ all the nitrogen is essentially in the form of N$^{4+}$ (the highest ionization state included in our models).


We expect the fine-structure far-infrared lines from N$^+$, O$^0$, and O$^{2+}$ to come primarily from the WIM or  \hii regions because their abundances will be much smaller in neutral clouds because of the rapid charge exchange and ion molecule reactions of N$^+$ and O$^+$ with H and H$_2$. In the WIM, if the dominant processes are proton charge exchange, X-ray photoionization, and electron radiative and di-electronic recombination,  then the distribution of lower ionized states of nitrogen and oxygen in the WIM will  be shifted towards higher ionization states for large X-ray fluxes.  We expect that the emission from lines such as  
[O\,{\sc iii}]\space at 52 \micron and 88 \micronno,   [N\,{\sc ii}]\space at 122 \micron and 205 \micronno, and [N\,{\sc iii}]\space at 57 \micronno, will also be suppressed in the presence of a strong X-ray flux, as is observed in a number of starburst and active galactic sources.



\section{Discussion}
\label{sec:discussion}

The solutions in Section~\ref{sec:results} show that it is possible for a high X-ray flux in galactic nuclei to alter the carbon ionization balance and reduce the C$^+$ abundance and correspondingly the \cii luminosity.  In the presence of a high flux of soft X-rays above 1 keV, a condition encountered in many active galactic nuclei \citep{Stacey2010,Ebrero2009}, the abundance of singly ionized carbon is reduced and converted into higher ionization states.   This reduction occurs primarily in the hot highly ionized gas that fills most of the galactic central zone, to some degree in the dense ionized skins surrounding  molecular clouds, less so in diffuse atomic hydrogen clouds, and very little, if at all, in the dense PDRs at the edge of the CO molecular cores.  Furthermore, in galactic nuclei with a high X-ray flux we would expect to see an increase in the dust temperature and infrared luminosity \citep{Voit1991b}.  Therefore in galactic nuclei, and in particular in active galactic nuclei, we expect a reduction in \cii emission relative to FIR/IR emission depending on the relative contribution of different ISM components to the \cii luminosity. Unfortunately the next most abundant ion, C$^{2+}$, does not have fine structure far-IR emission lines as its 2s$^2$ ground state has spin zero, and C$^{3+}$, although it has spin angular momentum due to its unpaired electron in the 2S level, does not have a nuclear spin or orbital angular momentum to break the degeneracy of the two electron spin states.  Instead to test the effect of X-ray ionization on the carbon balance we would need studies of their UV emission. For example the [C\,{\sc iv}] UV resonance lines are detected in extragalactic sources and have been used to trace the formation rate of massive stars \citep{Leitherer1991,Robert1993}.  [C\,{\sc iv}] UV absorption lines have also been used to study the properties of the Galactic Halo \citep{Savage2000} where carbon is presumed to be collisionally ionized in hot coronal gas \citep{Gnat2007}. 
  
To illustrate the potential effects of X-rays in reducing \cii emission we have run a number of simple models of \cii luminosity as a function of X-ray luminosity.  One model uses the ISM conditions of the CMZ and the other the higher densities appropriate to AGNs. To model any actual nucleus we would need to know many parameters including the properties of various ISM components -- their densities,  temperatures and filling factors, and the distribution and luminosities of the X-ray sources.  As will be seen below the primary effect of the X-rays is to reduce the \cii emission from the ionized gas.  Thus whether X-ray photoionization plays an important role or not depends on the relative contribution of the WIM and PDR components in galactic nuclei.  In the remainder of this section we present  simple models for the contribution to \cii emission from the ISM components in galactic nuclei and parameterize the effects of X-ray ionization on the reduction in \cii emission.  In one we will be guided by the properties of the ISM in the CMZ, but consider much larger X-ray luminosities than found there. In the other we consider much higher electron densities for the WIM and \h2 densities in PDRs, as well as a larger volume.  Our intent is not to model any particular galactic nucleus but to highlight the general behavior of the \cii emissivity in AGNs and similar active nuclei as a function of X-ray luminosity. 

\subsection{\cii emissivity from X-ray dominated nuclei}

We can estimate the \cii luminosity by summing the contributions of the different ISM components over their respective volumes.  To simplify our model calculations we assume that the physical parameters within each of these components are uniform and represent typical gas conditions.    The total galactic luminosity is

\begin{equation}
L_G(\mbox{\ciis}) = V_G \sum _{j}\eta_j \epsilon_j (\mbox{\ciis})\,\,{\rm erg\,\,s^{-1}}
\end{equation} 

\noindent where $V_G$ is the total volume of all \cii emitting regions, $\eta_j$ is the fractional volume occupied by each ISM component $j$, and $\epsilon_j$(\ciino) is the \cii emissivity in erg cm$^{-3}$ s$^{-1}$ of component $j$.  The emissivity for optically thin lines,

\begin{equation}
\epsilon_j (\mbox{\ciis}) = h\nu_{3/2,1/2} A_{3/2,1/2} n_{3/2,j}({\rm C^+})\,\, ({\rm erg\,\,cm^{-3}\,\,s^{-1}})
\end{equation}

\noindent where $A_{3/2,1/2}$ is the Einstein spontaneous radiation coefficient, $\nu_{3/2/,1/2}$ is the transition frequency, and $n_{3/2,j}$(C$^+$) is the number density of the upper $^2P_{3/2}$ state in ISM component $j$. In the optically thin limit one can solve exactly for the density of the $^2P_{3/2}$ state as a function of  the density and temperature of the collision partner \cite[see][]{Goldsmith2012,Langer2014},  

 \begin{equation}
n_{3/2,j}(C^+) =\frac{2n_j(X) n_t({\rm C^+})e^{-\Delta E/T_{\rm kin}}}{n_{j,{\rm cr}}(X)+n_j(X)(1+2e^{-\Delta E/T_{\rm kin}})}
\label{eqn:CII_excitation}
 \end{equation}

\noindent where $\Delta E$ =91.25K, $n_j(X)$ is the density of the dominant collision partner $X$ (= e, H, H$_2$) in each ISM component $j$, and $n_{j,{\rm cr}}(X)$ is the corresponding critical density \cite[see][]{Goldsmith2012,Wiesenfeld2014}.   In this paper we use the exact solution, but note that in the limit where the density of the collision partner, $n(X)$, is much less than the critical density, $n_{\rm cr}(X)$, Equation~\ref{eqn:CII_excitation} can be simplified to $n_{3/2}(C^+) \simeq 2 (n(X)/n_{\rm cr}(X))e^{-91.25/T_{\rm kin}}$. 
 

 
 \noindent  We define the average emissivity as $<$$\epsilon$(\ciino)$>$ = $L_G$(\ciino)/$V_G$, which upon substituting the line parameters for \cii yields,
 
  
  \begin{equation}
<\epsilon(\mbox{\ciis})> = 3.1\times10^{-20} \sum _{j} \eta_j f_j p_{3/2,j}({\rm C^+}) x_{{\rm tot},j}({\rm C}) n_j(X)
   \label{eqn:emissivity}
  \end{equation}

 \noindent where $f_j$ is the fractional abundance of C$^+$ with respect to all the charge states of ionized carbon for ISM component $j$, $p_{3/2,j}({\rm C^+}) = n_{3/2,j}({\rm C^+})/ n_t({\rm C^+})$ is the fractional population of the $^2P_{3/2}$ state, $x_{{\rm tot},j}$(C) is the total fractional abundance of carbon in all charge states with respect to the total hydrogen density ($n_t$= $n$(H) + 2$n$(H$_2$) + $n$(H$^+$)), and $n_j(X)$ is the dominant collision partner in component $j$.  
  
\subsection{Galactic nucleus with CMZ model parameters}
\label{sec:model_parameters}

In this section we set up a model galactic nucleus to study the \cii emissivity as a function of X-ray luminosity based on ISM parameters of the CMZ.  
To model the \cii emissivity from galactic nuclei we need to know the filling factor, density, temperature, and fractional abundance of C$^+$ for each ISM component.  We will use the conditions of the ISM in the CMZ as a guide for many, but not all components, and will consider a much wider range of X-ray luminosity than is found there.  The dominant ISM components in the nucleus are: 1) a dense warm ionized medium (WIM), 2) a low density hot ionized medium (HIM), 3) cold diffuse atomic hydrogen clouds (CNM), 4) a dense ionized skin (DIS) surrounding molecular clouds, 5) the dense PDRs of GMCs, and 6) compact \hii regions. We will neglect the contribution from the hot ionized medium because its densities are too low, $n$(e) $\sim$ few$\times$10$^{-2}$ cm$^{-3}$  and the compact \hii regions which, while bright, have too small a filling factor, to contribute to the bulk of the \cii emission. We consider the galactic nucleus to be a disk with a size similar to the CMZ.  We assume a radius $r$ = 200 pc which is the boundary of massive giant molecular clouds at $l \sim$ 358\fdg 7 in the CMZ as traced by CO \citep{oka1998a} and a scale height $z$ = 30 pc (disk thickness of 60 pc) slightly larger than the scale height  of the dense WIM \citep{Ferriere2007}.\footnote{The scale height of the WIM adopted for the CMZ is less than that derived from \cii for the Galactic disk where the density is much lower \citep{Langer2014z,Velusamy2014}.}

We adopt the following densities and temperatures for the key ISM components that contribute to the \cii emission from galactic nuclei: 1) the low density warm ionized medium is assumed to have $T_{\rm kin}$ = 8000K and an average density $n$(e) $\sim$ 10 cm$^{-3}$ 
\citep{Cordes2003,Roy2013}; 2) diffuse atomic hydrogen clouds with $n$(H) = 100 cm$^{-3}$ and $T_{\rm kin}$ = 100K, 3) dense PDRs with $n$(H$_2$) = 10$^3$ cm$^{-3}$ and $T_{\rm kin}$ = 100K, and for the dense ionized skins we adopt $n$(e) = 20 cm$^{-3}$ from the study of two clouds at the edge of the CMZ \citep{Langer2015}. These values are summarized in Table~\ref{tab:Table4}. 
We also assume a fractional abundance of carbon appropriate to the inner Galaxy, $x_{{\rm tot},j}$(C) = 5$\times$10$^{-4}$, which is an extrapolation of the metal abundance profile in the disk to inside 3 kpc \cite[see][]{Wolfire2003,Pineda2013}. 

We estimate the filling factor, $\eta$, of each component from the mass and density of the components. The mass of the GMCs in the CMZ is estimated to lie in the range (2 - 6)$\times$10$^7$ \Mss \citep{oka1998a,Ferriere2007} and we adopt the intermediate value 4$\times$10$^7$ \Ms. Their average H$_2$ densities are in the range 10$^4$ - 10$^5$ cm$^{-3}$.   We estimate the volume filled by the GMCs by dividing the total mass in GMCs,  by the density. For $n$(H$_2$) $\sim$ 10$^4$ cm$^{-3}$, the volume filled by clouds is $\sim$8$\times$10$^4$ pc$^3$, which yields a filling factor $\eta_{\rm H_2}$ about 1.1\% of the volume of the CMZ, for 10$^5$ cm$^{-3}$ the volume is only $\sim$8$\times$10$^3$ pc$^3$, or $\eta_{\rm H_2}$ $\sim$ 0.11\% of the CMZ volume.  The \cii emission from 
the dense PDR layer outside the CO cloud can be estimated from models of massive GMCs \citep{Wolfire2010} who find that the PDR layers lie in the range 1 to 1.5 magnitudes thick regardless of the cloud mass, UV radiation field, or hydrogen density.  Therefore the fractional volume filled by the PDR is just the difference between the volume occupied by the cloud out to the edge of the PDR minus that of the dense CO core.  For example, assuming a typical cloud mass of 10$^5$ \Mss and a core density of 10$^4$ cm$^{-3}$ and PDR density of 10$^{3}$ cm$^{-3}$, the PDR occupies $\sim$30 percent of the total cloud volume, which corresponds to $\sim$3$\times$10$^{-3}$ of the volume of the CMZ. For more massive clouds, 10$^6$ \Ms, and interior and PDR densities, 10$^5$ cm$^{-3}$ and 10$^4$ cm$^{-3}$, respectively, the PDRs occupy only $\sim$7$\times$10$^{-4}$ of the CMZ.  

Here we will consider a range of PDR filling factors, $\eta$(PDR) =  1$\times$10$^{-3}$ to 3$\times$10$^{-3}$. The choice of PDR filling factor has an influence on the models for two reasons, first, as shown above, it takes a very large X-ray flux to alter the carbon ionization balance in H$_2$ gas, and second, if the X-ray flux is large enough to alter the carbon ion abundance it is also large enough to destroy H$_2$.  Thus increasing the X-ray flux only pushes the PDR-like layer deeper into the cloud, but does not destroy it.  To simulate this situation in our model we assume that X-ray photoionization does not alter the C$^+$ abundance in the PDR component. 

The dense ionized skins occupy about 30\% of the volume of the CO core of the GMCs, for a filling factor $\eta$(DIS) $\sim$2$\times$10$^{-3}$, if the size of this region is similar to that derived for clouds in the region of {\it Sgr E}  \citep{Langer2015}. The diffuse atomic clouds have an estimated mass $\sim$5$\times$10$^5$ \Mss \citep{Ferriere2007} which corresponds to a filling factor $\sim$0.03 at a density $n$(H) = 100 cm$^{-3}$.  The ionized gas fills roughly 95\% of the volume of the galactic nucleus, of which 80\% is the dense WIM and $\sim$15\% the hot ionized medium (HIM) \cite[see][]{Ferriere2007} with negligible \cii emission.   

\begin{table}[htbp]																	
\caption{Galactic nucleus model parameters}
\label{tab:Table4}															
		\begin{tabular}{lccccccc}																	
\hline	
ISM  &  n(e)  & n(H) & n(H$_2$ )& $T_{\rm kin}$ & $\eta^a$ & Mass & \\
Component  & cm$^{-3}$ & cm$^{-3}$ &  cm$^{-3}$  &  K &  & \Ms \\
  \hline
\hline
HIM & 0.01  & 0 & 0 & 10$^5$ &  0.15  & 3$\times$10$^2$ \\
WIM & 10  & 0 & 0 & 8000 &  0.80  & 10$^6$  \\
\hi clouds  & 0.1 & 100 & 0 & 100 & 0.03 & 5$\times$10$^5$  \\
CO clouds & 0 & 0 & 10$^4$ & 50& 0.0125 & 6$\times$10$^7$ \\
PDR   & 3.5 & 0 & 10$^3$ & 100 & 0.002$^b$ &  2$\times$10$^6$ \\
DIS & 20  & 0 & 0 & 8000 &  0.003 & 10$^4$ \\
\hline
\end{tabular}
\\	
a.  The volume filling factor $\eta$ is dimensionless, and the sum for the various does not add up exactly to one because of roundoff. b. We  consider a range of values from 0.001 to 0.003, with 0.002 in the Table representing an intermediate value.
\end{table}

\subsubsection{\cii luminosity in the central X-ray model}

\begin{figure}
 \centering
      \includegraphics[width=9cm]{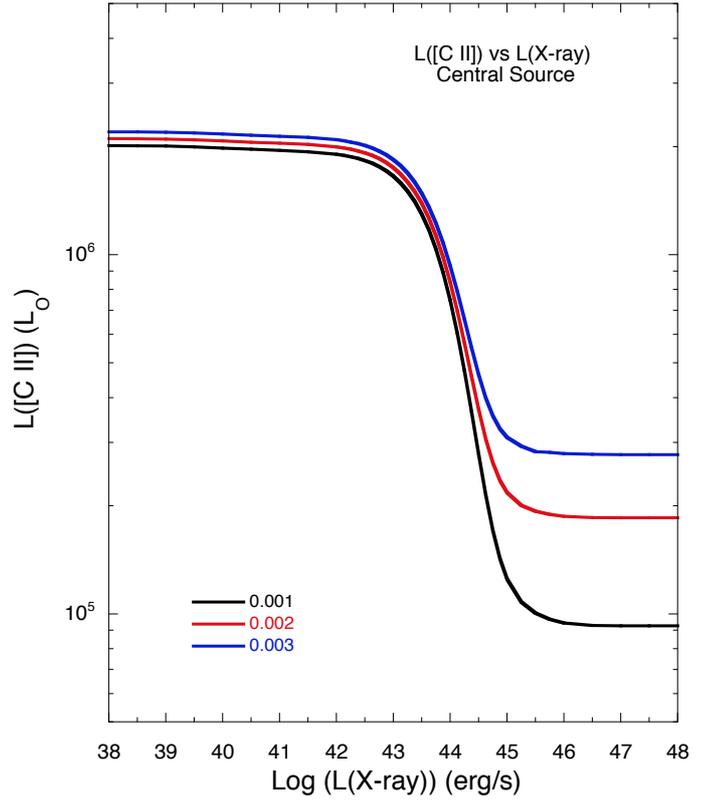}
         \caption{The \cii luminosity in units of solar luminosity as a function of the luminosity of a central X-ray source.  The assumed ISM parameters and dimensions of the nucleus are given in the text.  The three curves represent different values of the filling factor for PDR and PDR-like conditions in the nucleus.}
       \label{fig:fig6}
       \end{figure}

To demonstrate the effects of X-ray photoionization on the \cii emission we calculate the \cii luminosity as a function of X-ray luminosity for a central X-ray source, such as might be generated by a massive accreting black hole  using the ISM parameters in Table~\ref{tab:Table4}.  We calculate the X-ray flux as a function of radius assuming no attenuation and $r^{-2}$.  Fig.~\ref{fig:fig6} shows the total \cii luminosity, $L$(\ciis) in units of solar luminosity, $L_\odot$, as a function of X-ray luminosity,  $L$(X-ray), ranging from 10$^{38}$ to 10$^{48}$ erg s$^{-1}$ (the upper limit is a factor of a few beyond what is observed in AGNs).  Results are shown for three different choices of filling factor $\eta$(PDR) = 1, 2, and 3$\times$10$^{-3}$, and it can be seen that it takes an X-ray luminosity $\ge$10$^{42}$ erg s$^{-1}$ to have a measurable effect on the  C$^+$ balance and decrease the \cii luminosity in the galactic nucleus, and $>$10$^{43}$ erg s$^{-1}$ to have a significant effect. At the highest X-ray luminosities it is possible to suppress, by an order of magnitude or more, the \cii emission from the X-ray region of an AGN.  The asymptotic value is the contribution mainly from the PDRs. 

\begin{figure}
 \centering
                    \includegraphics[width=7.0cm]{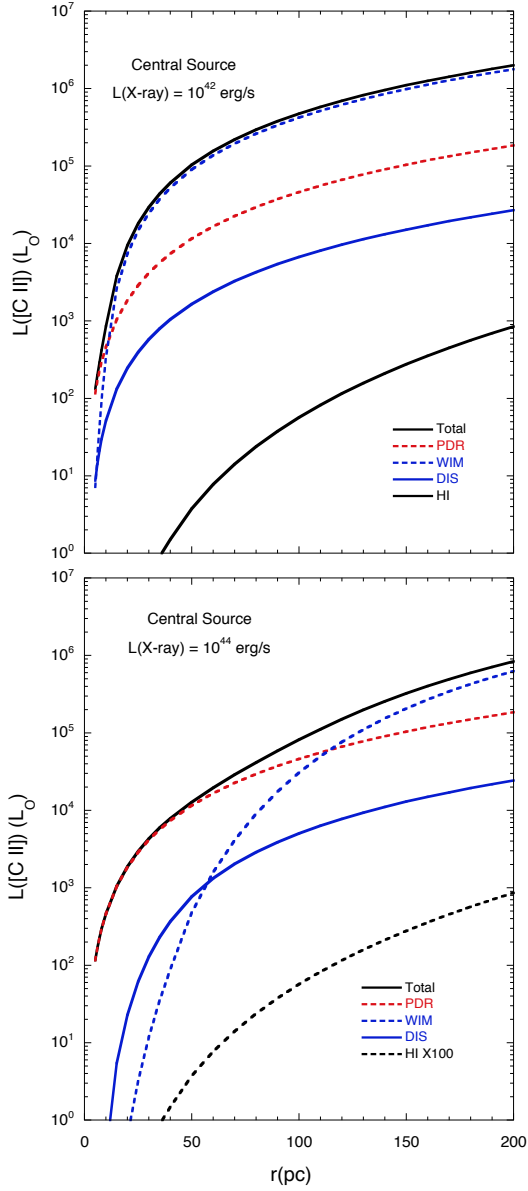}
      \caption{The \cii luminosities of the individual ISM components are plotted as a function of radius in a galactic nucleus with a central X-ray source with $L$(X-ray) = 10$^{42}$ (top panel) and 10$^{44}$ erg s$^{-1}$ (bottom panel) with a filling factor $\eta$(PDR) = 2$\times$10$^{-3}$.  The assumed ISM parameters and dimensions of the nucleus are given in the text.  The dense WIM dominates the \cii emission at $L$(X-ray) =10$^{40}$ erg s$^{-1}$ but is increasingly suppressed with increasing $L$(X-ray) until the PDRs dominate the emission.}
              \label{fig:fig7}
         \end{figure}

For a central X-ray source the contributions to the total luminosity will change as a function of galactic radius.  In Fig.~\ref{fig:fig7} we plot the contribution of the individual ISM components to $L$(\ciis) as a function of distance from the central source, $r$(pc), for two values of the central X-ray luminosity, $L$(X-ray) = 10$^{42}$ and 10$^{44}$ erg s$^{-1}$, adopting the intermediate filling factor $\eta$(PDR) = 2$\times$10$^{-3}$.   We chose these luminosities as examples because it is in this range of $L$(X-ray) that the photoionization of the WIM begins to make a difference in the relative contributions of the PDRs and the WIM.  For $L$(X-ray) = 10$^{42}$ erg s$^{-1}$ (Fig.~\ref{fig:fig7} top panel) the WIM dominates the contribution to \cii throughout the nucleus, followed by emission from the dense PDRs, while the diffuse \hi clouds and dense ionized skins (DIS) around the GMCs make negligible contributions to $L$(\ciis).  
At higher luminosities the X-rays have a more profound effect on the total emission, as shown in Fig.~\ref{fig:fig7} (bottom panel), and there is a different mix of ISM components contributing to $L$(\ciino). For $L$(X-ray) = 10$^{44}$ erg s$^{-1}$ PDRs dominate the \cii luminosity for $r \lesssim$ 100 pc, while the WIM dominates \cii luminosity outside 100 pc, and the \hi and DIS components are, again, negligible. 
Overall, the total \cii luminosity decreases with increasing $L$(X-ray) as was shown in Fig.~\ref{fig:fig6}. The ISM physical conditions in the WIM, DIS, dense PDRs, and atomic hydrogen clouds may be different in the AGNs, LIRGs, or ULIRGs from those found in the CMZ,  but the general dependence of the \cii luminosity on X-ray luminosity  will be similar. 



\subsubsection{\cii luminosity in the uniform X-ray flux model}

\begin{figure}
 \centering
 \includegraphics[width=8cm]{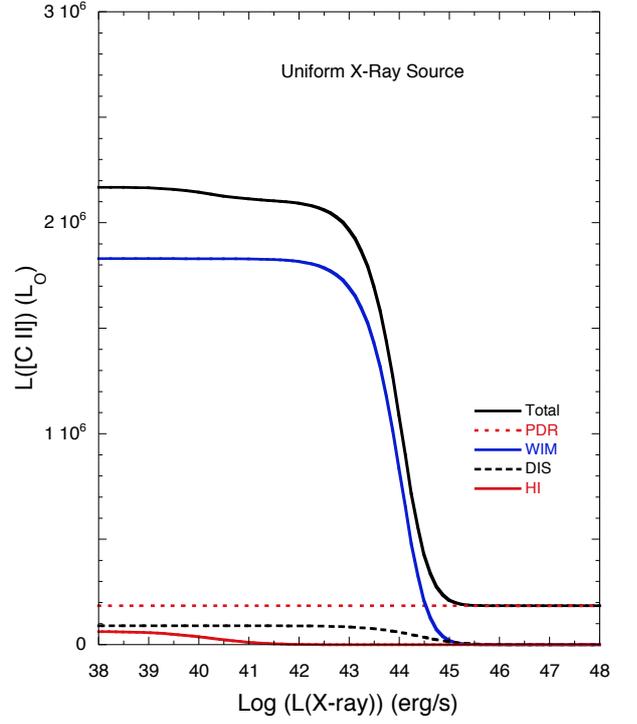}
      \caption{The luminosities of the individual ISM components  arising from a galactic nucleus with a uniform X-ray source, and physical parameters given in the text and Table~\ref{tab:Table4} are plotted as a function of X-ray luminosity. }
       \label{fig:fig8}
       \end{figure}

\begin{figure}
 \centering
 \includegraphics[width=7cm]{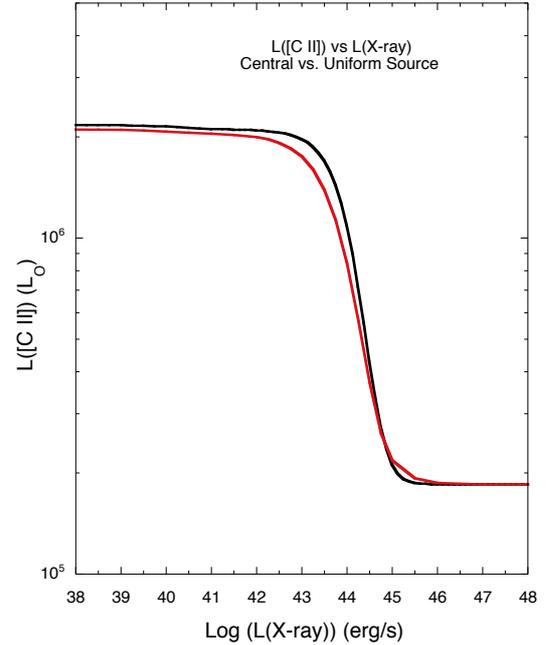}
      \caption{The \cii luminosity for the central X-ray source (black line) is compared to that of the uniform flux model (red line). The ISM properties are the same in both models.} 
       \label{fig:fig9}
       \end{figure}

Whereas in some galactic nuclei the X-ray luminosity is dominated by a central source, in others the luminosity may result from diffuse emission and/or numerous discrete sources spread throughout.  To model this situation we assume a uniform flux throughout the nucleus and calculate the luminosity, $L$(X-ray) = $F_{X} A$, where A is the  area of the boundary of the galactic nucleus.  We adopt the same disk model with a radius of 200 pc and 60 pc thick used in the CMZ model with a central X-ray source.  The exact dimensions are not critical because the results can be extended to galaxies with X-ray regions of different size through the relationship between flux, emissivity, and luminosity for a uniform environment.  The ISM parameters are again given in Table~\ref{tab:Table4} and we adopt the intermediate value $\eta$(PDR) = 2$\times$10$^{-3}$. The X-ray luminosity for a uniform flux is just proportional to the area of the outermost volume, $L$(X-ray)=$F_X$A$_{\rm AGN}$, where A$_{\rm AGN}$ is the area of the AGN's uniform X-ray source, while the \cii luminosity is proportional to the emissivity times the volume, $L$(\ciis) = $\epsilon$(\ciis)$V_{\rm AGN}$, where  $V_{\rm AGN}$ is the corresponding volume of the uniform X-ray source. Therefore, the results plotted below using the dimensions of the CMZ can be converted to different AGN volumes as follows,

\begin{equation}
\begin{array} {l}
L_{AGN}(\mbox{\ciis}) = (V_{AGN}/V_{CMZ}) L_{CMZ}(\mbox{\ciis}) \\ \\
L_{AGN}(\mbox{X-ray}) = (A_{AGN}/A_{CMZ}) L_{CMZ}(\mbox{X-ray}),
\end{array}
\end{equation}

\noindent in which one scales a given AGN X-ray luminosity by the ratio of areas to the CMZ and then determines the CMZ \cii luminosity and scales it by the ratio of the volumes to derive the AGN \cii luminosity.
 In Fig.~\ref{fig:fig8} we plot the \cii luminosity arising from a galactic nucleus with uniform X-ray source.  The luminosities of the individual ISM components are plotted as a function of X-ray luminosity. The dense warm ionized medium dominates the \cii luminosity up to $L$(X-ray) $\sim$10$^{44}$ erg s$^{-1}$, after which PDRs, with a filling factor $\eta$(PDR) = 2$\times$10$^{-3}$,  dominate the emission.  
Finally, in Fig.~\ref{fig:fig9}  we compare the \cii luminosity of the central and uniform X-ray models as a function of $L$(X-ray) assuming the same CMZ ISM conditions and $n$(e)=10 cm$^{-3}$ for the WIM and the same value for $\eta$(PDR) = 2$\times$10$^{-3}$.  The two solutions for $L$(\ciis)  are very similar.  At low X-ray luminosity the flux is too low to influence the C$^+$ abundance and the emissivities are independent of $L$(X-ray), while high luminosities the C$^+$ abundances are reduced in the WIM throughout the volume and the PDRs dominate the emission.  At intermediate X-ray  luminosities the solutions for the central source reduces the C$^+$ abundance primarily in the interior (smaller $r$) while having less of an effect near the outer part of the nucleus, and this tends to average out to a solution close to that of having a uniform flux. 

             
\subsection{AGN representative model of \cii emission}

In the CMZ model we adopted an average electron density $n$(e) = 10 cm$^{-3}$, but the most central region has been estimated to have a much higher density \cite[c.f.][]{Ferriere2007}.  For example, \cite{Mezger1979} used thermal radio continuum radio observations to derive a central $n$(e) $\sim$ 26 cm$^{-3}$ while \cite{Mehringer1992,Mehringer1993} derive even higher electron densities outside compact HII regions near Sgr B1 and B2 and derive an average electron density of $\sim$ 60 and 80 cm$^{-3}$ over two regions $\sim$ 700 and 170 pc$^2$, respectively.  The conditions encountered in the most energetic region of the CMZ is likely to prevail over a much larger volume in AGNs.  In addition, the \niii luminosity in some galactic nuclei is comparable to that of \nii but requires higher electron densities to excite its $^2P_{3/2}$ level. Thus it is reasonable to consider a model for AGNs with much higher WIM densities than the CMZ. In addition, the PDR densities are also likely higher \citep{Stacey2010}. To illustrate the impact of higher electron densities we first demonstrate the sensitivity of the WIM \cii luminosity as a function of $n$(e) on a stand alone basis and then in a simple model of AGNs.

\subsubsection{\cii luminosity from the WIM}

\begin{figure}
 \centering
 \includegraphics[width=7cm]{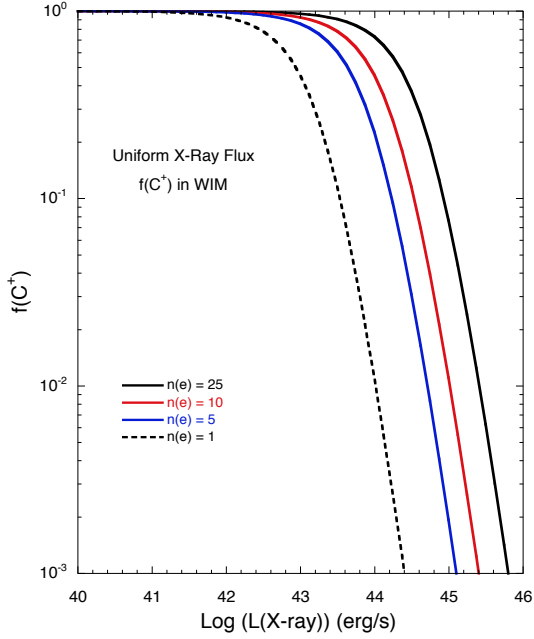}
      \caption{The fraction of singly ionized carbon in the WIM versus X-ray luminosity in the uniform X-ray flux model, for a range of WIM electron densities.}
       \label{fig:fig10}
       \end{figure}

\begin{figure}
 \centering
 \includegraphics[width=7cm]{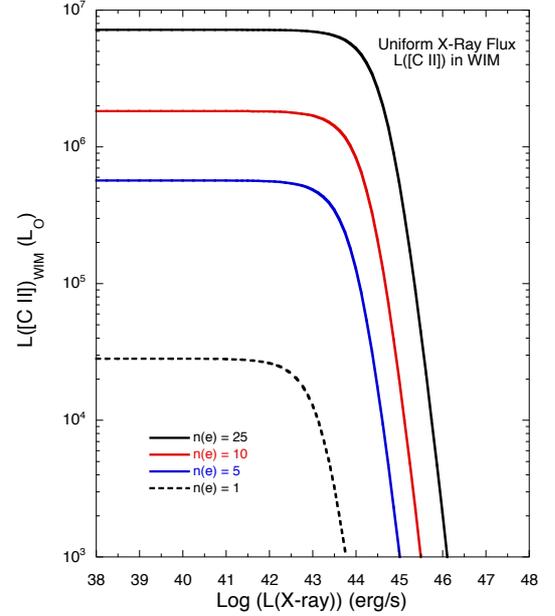}
      \caption{The \cii luminosity arising from the WIM  versus X-ray luminosity, for a range of electron densities in the WIM. The remaining ISM parameters are the same as that for the uniform model.}
       \label{fig:fig11}
       \end{figure}

\begin{figure}
 \centering
 \includegraphics[width=7cm]{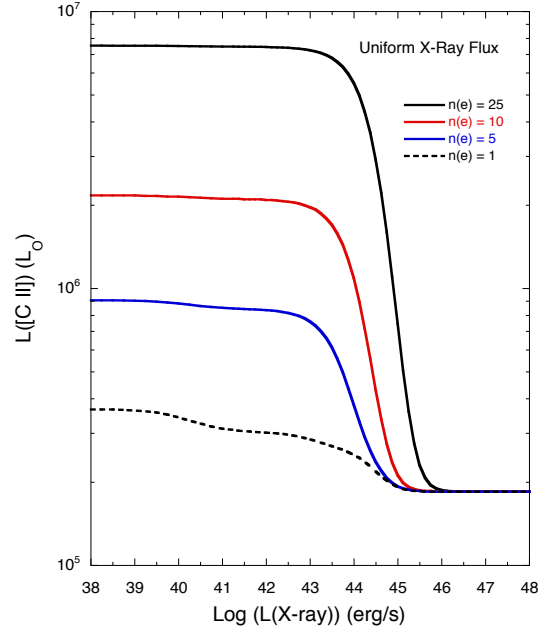}
      \caption{The total \cii luminosity (all ISM components) versus  X-ray luminosity, for a range of electron densities in the WIM. The remaining ISM parameters are the same as that for the uniform model.}
       \label{fig:fig12}
       \end{figure}

In the model adopted here for the \cii luminosity from active galactic nuclei the WIM dominates the emission of \cii over a wide range of X-ray luminosities.  However, the model adopts a fixed electron density in the WIM, $n$(e) = 10 cm$^{-3}$.  To understand better the sensitivity to the choice of this parameter we have also calculated the luminosity from the WIM, $L_{\rm WIM}$(\ciis), versus  the electron density. Fig.~\ref{fig:fig10} shows the fraction of singly ionized carbon in the WIM as a function of X-ray luminosity for a range of WIM electron abundances, $n$(e) = 1 to 25 cm$^{-3}$.  As expected, the smaller the electron abundance the more profound is the effect of X-rays on $f$(C$^+$) for a given X-ray luminosity, because the electron recombination rate of C$^+$ cannot as efficiently offset the X-ray photoionization rate. In the absence of X-rays the luminosity from the WIM would be expected to decrease roughly like $n$(e)$^{-2}$ below the electron critical density ($n_{cr}$(e)$\sim$ 40 cm$^{-3}$). However, we expect that the \cii luminosity from the WIM decreases faster than the density $n$(e) in the presence of a large X-ray flux because not only does the collisional excitation rate decrease, but so does the fraction of available C$^+$.  This nonlinear effect can be seen in Fig.~\ref{fig:fig11}, which plots $L_{\rm WIM}$ as a function of $L$(X-ray) for the range of $n$(e) plotted  in Fig.~\ref{fig:fig10}.  In Fig.~\ref{fig:fig12} we plot the total \cii luminosity, $L_{\rm tot}$ as a function of $L$(X-ray) for a range of $n$(e) in the WIM, and it can be seen that the drop in \cii luminosity can be quite large if the WIM electron density is higher, such as might be expected in active galactic nuclei.

       
\subsubsection{\cii luminosity from an AGN}

\begin{figure}
 \centering
 \includegraphics[width=8.5cm]{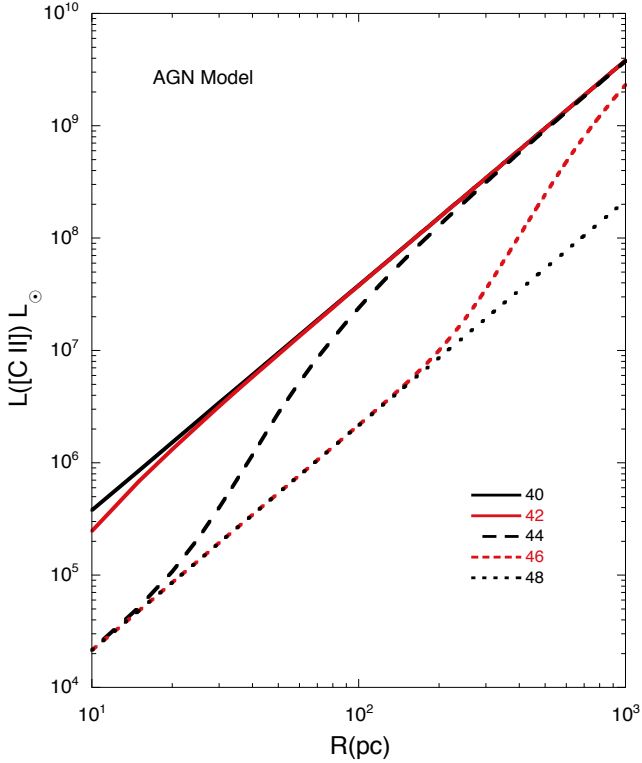}
      \caption{The total \cii luminosity of a hypothetical AGN  versus  galactic radius for a range of X-ray luminosities. The key labels the log of the X-ray luminosity and ranges from Log$L$(X-ray) = 40 to 48.  The decrease in \cii luminosity with increasing $L$(X-ray is due to the destruction of C$^+$ in the WIM close to the source of X-rays.  The radius over which the X-rays photoionize C$^+$ increases with increasing X-ray luminosity.}
       \label{fig:fig13}
       \end{figure}

We model a hypothetical AGN assuming a disk geometry with a central X-ray source.  The WIM electron density is $n$(e) = 50 cm$^{-3}$ 
and the PDR has $n$(H$_2$) = 5$\times$10$^3$ cm$^{-3}$.  The molecular mass in active galactic nuclei is larger than the CMZ but as the density of the giant molecular clouds is an order of magnitude higher the filling factor is not much different from what we calculated for the CMZ and we assume $\eta$=0.002.  One other difference is the scale height of the gas which is likely to be higher than in the CMZ due to a higher pressure and we adopt a scale height for CO clouds and the WIM of 100 pc and 200 pc, respectively.  The \cii luminosity for this AGN model is plotted in Fig.~\ref{fig:fig13} as a function of galactic radius from the central source for a range of $L$(X-ray). The X-ray flux close to the central accretion hole ionizes the C$^+$ and reduces the luminosity of the WIM as can be seen in the decrease in the luminosity as a function of X-ray luminosity.  As one goes further from the central source the WIM increases and contributes to the total.  At a few hundred pc there is roughly an order of magnitude difference in the total luminosity. The $L$(\ciino) dependence on  $L$(X-ray) is similar to the CMZ example except that the luminosity is larger due to higher electron and H$_2$ densities.

       
 \subsection{Observational test case}
 
\begin{figure}
 \centering
  \includegraphics[width=8.5cm]{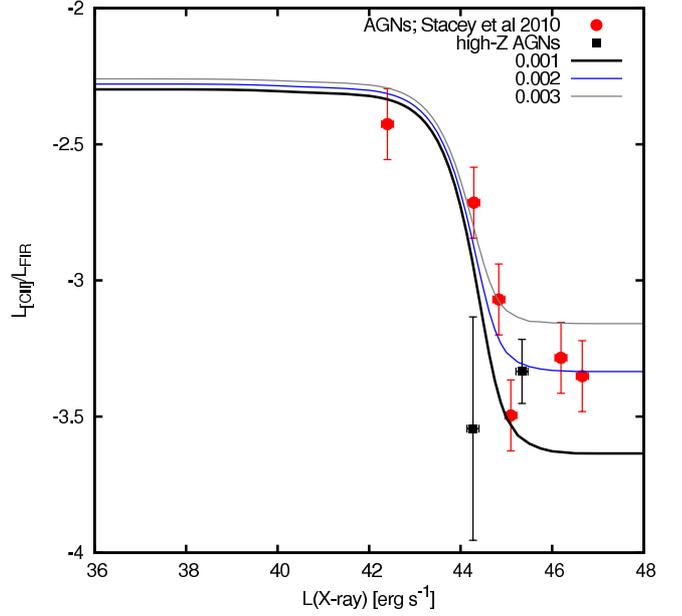}
      \caption{The predicted [C\,{\sc ii}]/FIR luminosity ratio versus X-ray luminosity for different PDR filling factors (see Fig.~\ref{fig:fig6}).  
      Also plotted are data points for six AGNs (red bullets) observed in \ciino, FIR, and X-rays \citep{Stacey2010} and  two  high-redshift AGNs (black squares) observed in \cii (as compiled in \citet{Gullberg2015}), FIR by \citet{Vignali2005},  and X-rays by \citet{Law2004}. We assumed 30\% uncertainties for the [C\,{\sc ii}]/FIR ratios from \citet{Stacey2010} and for the X--ray luminosities. } 
\label{fig:fig14}
       \end{figure}

In Fig.~\ref{fig:fig14}, we compare our model calculations to a sample of six extragalactic sources for which the \ciino, FIR, and X-ray luminosities have been measured (\citealt{Stacey2010} and  references therein). We also include two high-redshift AGNs compiled by \citet{Gullberg2015} in [C\,{\sc ii}] and FIR and by  \citet{Vignali2005} and \citet{Law2004} in X--rays. We calculated the model [C\,{\sc ii}]/FIR ratio by dividing the predicted [C\,{\sc ii}]
 luminosity for different PDR filling factors (Fig.~\ref{fig:fig6}) by the FIR luminosity of the CMZ, $L_{\rm FIR}=4\times10^{8}$ L$_{\odot}$ \citep{Launhardt2002}.  The observed X-ray luminosities cover 2 to 10 keV and, because the  spectral energy distribution increases at lower energies  the actual luminosity would be a factor $\sim$3.8 larger for E$>$ 1 keV assuming the spectral energy distribution in Equation~\ref{eqn:dJdE} with $\Gamma$ = 1.9. Thus the data points would shift slightly to the right by $\sim$0.59 on the log scale, which is still consistent with our model calculation.  We see that for this sample of AGNs the [C\,{\sc ii}]/FIR luminosity ratio is in good agreement with our model predictions, showing significantly lower [C\,{\sc ii}]/FIR ratios for $L$(X-ray)$>$10$^{43}\,{\rm erg}\,{\rm s}^{-1}$.


\subsection{Central Molecular Zone \cii luminosity}

The Galactic Central Molecular Zone provides the best opportunity to check our model of \cii luminosity in the presence of X-rays because we have the best information about the state of its ISM components, at least as compared to external galaxies. However, in practice the uncertainties in its ISM conditions are too large to allow more than an estimate of the  \cii luminosity.  

Unfortunately, the Galactic Central Molecular Zone is not a strong source of X-ray luminosity.  {\it Chandra} has detected $\gtrsim$9000 discrete stellar sources in the innermost CMZ \citep{Muno2009}, but most of these have luminosities in the range $L$(X-ray) = 10$^{31}$ and 10$^{33}$ erg s$^{-1}$ with a few having maximum values $\sim$ few$\times$10$^{34}$ erg s$^{-1}$.  Adopting an average luminosity per source of 10$^{32}$ erg s$^{-1}$ yields a maximum extended luminosity of order 10$^{36}$ erg s$^{-1}$, which is too low to have much of an effect on the \cii luminosity based on the results in Fig.~\ref{fig:fig8}. The strongest point source in the CMZ, 1E 1740.7--2942, is believed to be an accreting stellar black hole and has a luminosity, $L$(X-ray $\sim$ few$\times$10$^{37}$ erg s$^{-1}$ \citep[see][]{Maloney1997}.  The black hole assumed associated with {\it Sgr A} at the Galactic Center has a current luminosity $\sim$10$^{33}$ erg s$^{-1}$ \citep{Baganoff2001}, but may have been brighter in the recent past.  Thus X-rays likely have little effect on the \cii luminosity of the CMZ. and so it is not a good source for testing the effects of X-rays on \cii luminosity.  Nonetheless, the CMZ \cii luminosity does provide an opportunity to validate whether our  model predicts the right order of magnitude for the \cii luminosity.

To calculate the \cii luminosity for the CMZ we use the CMZ physical dimensions adopted in Section~\ref{sec:model_parameters} and ISM parameters in Table~\ref{tab:Table4} for the atomic clouds, PDRs, and DIS components, but modify the WIM model to fit with the WIM electron profile \citep{Ferriere2007}, as discussed below.  The atomic, molecular, and ionized gas in the CMZ is concentrated at the center and decreases with radius and distance above and below the plane.  We are interested in calculating the \cii luminosity arising from the volume of the CMZ so the distribution of the atomic clouds and the PDRs and DISs associated with the giant molecular clouds is not important under the assumption that the clouds have the same characteristics throughout the CMZ volume, as assumed here,  and as long as the X-ray flux has a minimal impact on the C$^+$ abundance.  The WIM density, in contrast, does change throughout the  CMZ.  \cite{Ferriere2007} summarizes the results from  \cite{Lazio1998} and \cite{Cordes2002,Cordes2003} and expresses the electron space averaged density as the sum of three terms, one of which, $<n$(e)$>_3(r,z)$, dominates in the CMZ with a profile that decreases as a Gaussian in $r$ and $z$ (the other two terms are smaller in absolute value but have much larger scale factors and so decrease very slowly in the CMZ). We have simplified this expression by neglecting the offset terms in   $<n$(e)$ >$ and setting $<n$(e) $>_1$ and $<n$(e)$ >_2$ constant in the CMZ, which yields

\begin{equation}
<n(\mbox{e})>(r,z) = 10e^{-(r^2/r_0^2)} e^{-(z^2/z_0^2)} + 0.14\,\, \mbox{cm$^{-3}$}
\end{equation}

\noindent where the peak density is $\sim$ 10 cm$^{-3}$ at the center and decreases with Gaussian parameters $r_0$ = 145 pc and $z_0$ = 26 pc.

We have calculated the \cii emissivity throughout the CMZ disk ($r \leq$ 200 pc and $|$z$|$ $\leq$ 30 pc) using the model of the CMZ discussed above and assuming a central source with X-ray luminosity 10$^{37}$ erg s$^{-1}$.  We integrate the emissivity over the volume of the CMZ to derive an emergent  \cii luminosity, $L$(\ciino) $\sim$ 7 - 9 $\times$10$^5$ $L_\odot$, for $\eta$(PDR) = (1 - 3)$\times$10$^{-3}$. This luminosity hardly differs from that without an X-ray source because, as shown earlier, the X-ray luminosity  is too low to modify the C$^+$ distribution in the ISM components. For comparison we have calculated the \cii  luminosity measured by BICE using the line intensity contours in Fig. 8 of \cite{Nakagawa1998} integrated over the area of the CMZ.  We estimate that BICE observes $L$(\ciino) $\sim$ 4.5$\times$10$^5$ erg s$^{-1}$.  Our $L$(\ciino) model of the CMZ and the BICE observations agree within a factor of 2 or better which is probably about the best that we can expect given the uncertainties in ISM parameters assumed in our model calculation. 
In summary, the reasonable agreement of our \cii emissivity model and the observed \cii luminosity of the CMZ lends confidence to the approach presented here.


\section{Summary}
\label{sec:summary}

We have explored the influence of X-rays on the ionization balance of carbon in the interstellar medium with a focus on its effect in galactic nuclei with large X-ray luminosity.  In the presence of a large X-ray flux we find that C$^+$ can be converted to higher ionization states,  C$^{j+}$ with $j\ge$ 2.  The effect of X-rays on the \cii luminosity of galactic nuclei depends on the X-ray luminosity and the details of the properties of the ISM, namely the density, temperature, and filling factors for each of the ISM components.  The biggest impact is on the ionization distribution in the WIM, where C$^+$ is more readily converted to C$^{2+}$, C$^{3+}$, etc., thus reducing its contribution to the \cii luminosity.  

Thus X-ray photoionization is another mechanism to consider in understanding the reduction of the  \cii to far-IR luminosity ratio  in active galactic nuclei.   It appears that, on average, it requires a soft X-ray luminosity greater than about 10$^{42}$ - 10$^{43}$ erg s$^{-1}$  to have a noticeable effect on the \cii luminosity.  Our Galaxy does not fall into this category, but many others do, in particular some Luminous Infrared Galaxies (LIRGs) and Ultra Luminous Infrared Galaxies (ULIRGs), and many Active Galactic Nuclei (AGNs).  
If the primary X-ray source is an accreting black hole then the \cii will be suppressed in its vicinity but not necessarily throughout the entire galactic nucleus. However, as discussed above, AGNs have, on average, considerably larger X-ray luminosities with $L$(X-ray) ranging up to a few $\times$10$^{47}$ erg s$^{-3}$ \citep{Ebrero2009}. In the AGNs with $L$(X-ray) $\gtrsim$ 10$^{43}$ erg s$^{-1}$, depending on the details of the ISM parameters, X-rays can suppress the \cii luminosity.  It is precisely these sources that consistently have a much lower \cii luminosity compared to their far-infrared dust luminosity  \citep[see summary by][]{Gullberg2015} which may be due, in part, to the influence of X-rays on the C$^+$ photoionization into higher ionization states.

We have also briefly considered the ionization balance of nitrogen and oxygen in the dense warm ionized medium under the influence of X-rays.  We find that the lower ionization states of nitrogen and oxygen are suppressed at high X-ray fluxes and at the highest flux values these elements are essentially in the highest ionization state.  We did not calculate detailed models of the emission  from the far-infrared lines of the lower ionization states, but it is apparent, by comparison with the results for carbon, then the emission from these tracers will be suppressed at high fluxes.  \niii is observed to be strong in some galaxies which requires an ionization source for N and N$^+$.  The usual explanation invokes an EUV flux, but X-rays are an alternative explanation and have the advantage of being present throughout galactic nuclei. Thus X-ray photoionization may explain some of the decrease in emission from carbon, nitrogen, and oxygen ions and the increase in emission of higher ionization states, such as \niii and [O\,{\sc iii}], in active galactic nuclei.

We have been guided by conditions in the Galactic Central Molecular Zone, but AGNs will have different ISM parameters of density, temperature, and filling factor for the various gas components and each one would need to be analyzed on a case by case basis.  For comparison we adopted a model of an AGN with higher WIM and PDR densities. We also compared our \cii luminosity models with a small data set of extrgalactic sources with high X-ray luminosities.  We find that their \cii to FIR ratio decreases significantly at an X-ray luminosity consistent with our model calculations. Finally, to confirm that the \cii emissivity model we use is reasonable we calculated the \cii luminosity for the CMZ where conditions are reasonably well known, at least in comparison to extragalactic sources.  The Galactic CMZ model  produces a \cii luminosity $\sim$  7 - 9 $\times$10$^5$ $L_\odot$, depending on the details of parameters chosen for the WIM and PDRs.  This \cii  luminosity compares reasonably well, given all the uncertainties in the ISM conditions,   to the \cii luminosity of $\sim$ 4.5$\times$10$^5$ $L_\odot$ we estimate from the BICE observations.  

In conclusion, X-ray photoionzation can alter the ionization balance of the ionization states of carbon, nitrogen, and oxygen primarily in the ionized ISM, resulting in a significant  fraction of their abundance shifting tohighly ionized states in active galactic nuclei. Therefore the effects of X-rays on the ionization balance needs to be included in any modeling and interpretation of carbon, nitrogen, and oxygen far-IR line emission in active galactic nuclei.  


\begin{acknowledgements}
We would like to thank an anonymous referee for a careful reading of our manuscript and several suggestions that improved the models and discussion. We would also like to thank Dr. P. F. Goldsmith for useful comments on the text and  Dr. B. Drouin for discussions regarding the spin states of [C\,{\sc iii}] and [C\,{\sc iv}].  This work was performed at the Jet Propulsion Laboratory, California Institute of Technology, under contract with the National Aeronautics and Space Administration.   {\copyright}2015 California Institute of Technology: USA Government sponsorship acknowledged.
 
\end{acknowledgements}


\bibliographystyle{aa}
\bibliography{aa_CII_X_ray_refs}


\end{document}